\documentclass[sigconf, nonacm]{acmart}
\pdfoutput=1

\newcommand\vldbdoi{XX.XX/XXX.XX}

\newcommand\vldbavailabilityurl{https://github.com/itrummer/DataCorrelationPredictionWithNLP}
\usepackage{tikz}
\usetikzlibrary{shadows.blur}
\usetikzlibrary{shapes.geometric}
\usepackage{pgfplots}
\usepgfplotslibrary{groupplots}
\usepackage{verbatim}
\usepackage{etoolbox}
\usepackage{caption}
\usepackage{subcaption}

\begin{document}
\title{Can Deep Neural Networks Predict Data Correlations from Column Names?}

%%
%% The "author" command and its associated commands are used to define the authors and their affiliations.
\author{Immanuel Trummer}
\affiliation{%
  \institution{Cornell Database Group}
  \city{Ithaca, NY}
  \state{USA}
}
\email{itrummer@cornell.edu}

\newtheorem{hypothesis}{Hypothesis}
\newtheorem{answer}{Answer}

\newcommand{\rev}[1]{\textcolor{black}{#1}}
\newcommand{\rew}[1]{\textcolor{blue}{#1}}

%\input{sections/replies}

%%
%% The abstract is a short summary of the work to be presented in the
%% article.
\begin{abstract}
Recent publications suggest using natural language analysis on database schema elements to guide tuning and profiling efforts. The underlying hypothesis is that state-of-the-art language processing methods, so-called language models, are able to extract information on data properties from schema text.

This paper examines that hypothesis in the context of data correlation analysis: is it possible to find column pairs with correlated data by analyzing their names via language models? First, the paper introduces a novel benchmark for data correlation analysis, created by analyzing thousands of Kaggle data sets (and available for download). Second, it uses that data to study the ability of language models to predict correlation, based on column names. The analysis covers different language models, various correlation metrics, and a multitude of accuracy metrics. It pinpoints factors that contribute to successful predictions, such as the length of column names as well as the ratio of words. Finally, \rev{the study analyzes the impact of column types on prediction performance.} The results show that schema text can be a useful source of information and inform future research efforts, targeted at NLP-enhanced database tuning and data profiling.
\end{abstract}
%the study evaluates a data profiling algorithm that uses natural language analysis to guide profiling efforts.

% recent advances in natural language processing enable automated systems to extract information on data properties from 

% powerful language models are able to infer 

% useful information for data profiling efforts. This paper studies the question whether this 

% For humans, it is often possible to predict data correlations from column names. We conduct experiments to find out whether deep neural networks can learn to do the same. If so, e.g., it would open up the possibility of tuning tools that use NLP analysis on schema elements to prioritize their efforts for correlation detection. 

% We analyze correlations for around 120,000 column pairs, taken from around 4,000 data sets. We try to predict correlations, based on column names alone. For predictions, we exploit pre-trained language models, based on the recently proposed Transformer architecture. We consider different types of correlations, multiple prediction methods, and various prediction scenarios. We study the impact of factors such as column name length or the amount of training data on prediction accuracy. Altogether, we find that deep neural networks can predict correlations with a relatively high accuracy in many scenarios (e.g., with an accuracy of 95\% for long column names). 

%complement existing correlation prediction methods.
\maketitle

%%% do not modify the following VLDB block %%
%%% VLDB block start %%%
% \pagestyle{\vldbpagestyle}
% \begingroup\small\noindent\raggedright\textbf{PVLDB Reference Format:}\\
% \vldbauthors. \vldbtitle. PVLDB, \vldbvolume(\vldbissue): \vldbpages, \vldbyear.\\
% \href{https://doi.org/\vldbdoi}{doi:\vldbdoi}
% \endgroup
% \begingroup
% \renewcommand\thefootnote{}\footnote{\noindent
% This work is licensed under the Creative Commons BY-NC-ND 4.0 International License. Visit \url{https://creativecommons.org/licenses/by-nc-nd/4.0/} to view a copy of this license. For any use beyond those covered by this license, obtain permission by emailing \href{mailto:info@vldb.org}{info@vldb.org}. Copyright is held by the owner/author(s). Publication rights licensed to the VLDB Endowment. \\
% \raggedright Proceedings of the VLDB Endowment, Vol. \vldbvolume, No. \vldbissue\ %
% ISSN 2150-8097. \\
% \href{https://doi.org/\vldbdoi}{doi:\vldbdoi} \\
% }\addtocounter{footnote}{-1}\endgroup
%%% VLDB block end %%%

%%% do not modify the following VLDB block %%
%%% VLDB block start %%%
\ifdefempty{\vldbavailabilityurl}{}{
\vspace{.3cm}
\begingroup\small\noindent\raggedright\textbf{Artifact Availability:}\\
The source code, data, and/or other artifacts have been made available at \url{\vldbavailabilityurl}.
\endgroup
}

%%% do not modify the following VLDB block %%
%%% VLDB block start %%%
% \pagestyle{\vldbpagestyle}
% \begingroup\small\noindent\raggedright\textbf{PVLDB Reference Format:}\\
% \vldbauthors. \vldbtitle. PVLDB, \vldbvolume(\vldbissue): \vldbpages, \vldbyear.\\
% \href{https://doi.org/\vldbdoi}{doi:\vldbdoi}
% \endgroup
% \begingroup
% \renewcommand\thefootnote{}\footnote{\noindent
% This work is licensed under the Creative Commons BY-NC-ND 4.0 International License. Visit \url{https://creativecommons.org/licenses/by-nc-nd/4.0/} to view a copy of this license. For any use beyond those covered by this license, obtain permission by emailing \href{mailto:info@vldb.org}{info@vldb.org}. Copyright is held by the owner/author(s). Publication rights licensed to the VLDB Endowment. \\
% \raggedright Proceedings of the VLDB Endowment, Vol. \vldbvolume, No. \vldbissue\ %
% ISSN 2150-8097. \\
% \href{https://doi.org/\vldbdoi}{doi:\vldbdoi} \\
% }\addtocounter{footnote}{-1}\endgroup
% %%% VLDB block end %%%

% %%% do not modify the following VLDB block %%
% %%% VLDB block start %%%
% \ifdefempty{\vldbavailabilityurl}{}{
% \vspace{.3cm}
% \begingroup\small\noindent\raggedright\textbf{PVLDB Artifact Availability:}\\
% The source code, data, and/or other artifacts have been made available at \url{\vldbavailabilityurl}.
% \endgroup
% }
% %%% VLDB block end %%%

\pgfplotsset{scaled y ticks=false, log ticks with fixed point}
\pgfplotsset{
    discard if/.style 2 args={
        x filter/.code={
            \edef\tempa{\thisrow{#1}}
            \edef\tempb{#2}
            \ifx\tempa\tempb
                \def\pgfmathresult{inf}
            \fi
        }
    },
    discard if not/.style 2 args={
        x filter/.code={
            \edef\tempa{\thisrow{#1}}
            \edef\tempb{#2}
            \ifx\tempa\tempb
            \else
                \def\pgfmathresult{inf}
            \fi
        }
    }
}

\newlength{\plotHeight}
\setlength{\plotHeight}{5.25cm}

\def\plotTwo#1#2{\begin{tikzpicture}
        \begin{groupplot}[group style={group size=5 by 1, y descriptions at=edge left, horizontal sep=5pt}, width=2.75cm, height=\plotHeight, ybar=0pt, ymajorgrids, ylabel=Score, ylabel near ticks, ymin=0, ymax=1, xtick=\empty, legend entries={#2}, legend columns=4, legend to name=lastLegend, nodes near coords, every node near coord/.append style={anchor=east, rotate=90}]
            \nextgroupplot[title=F1]
            \addplot+[discard if not={id}{1}] table[col sep=comma, x expr=0, y=f1] {#1};
            \addplot+[discard if not={id}{2}] table[col sep=comma, x expr=0, y=f1] {#1};
            \nextgroupplot[title=Recall]
            \addplot+[discard if not={id}{1}] table[col sep=comma, x expr=0, y=rec] {#1};
            \addplot+[discard if not={id}{2}] table[col sep=comma, x expr=0, y=rec] {#1};
            \nextgroupplot[title=Precision]
            \addplot+[discard if not={id}{1}] table[col sep=comma, x expr=0, y=pre] {#1};
            \addplot+[discard if not={id}{2}] table[col sep=comma, x expr=0, y=pre] {#1};
            \nextgroupplot[title=MCC]
            \addplot+[discard if not={id}{1}] table[col sep=comma, x expr=0, y=mcc] {#1};
            \addplot+[discard if not={id}{2}] table[col sep=comma, x expr=0, y=mcc] {#1};
            \nextgroupplot[title=Accuracy]
            \addplot+[discard if not={id}{1}] table[col sep=comma, x expr=0, y=acc] {#1};
            \addplot+[discard if not={id}{2}] table[col sep=comma, x expr=0, y=acc] {#1};
        \end{groupplot}
    \end{tikzpicture}}

\def\plotThree#1#2{\begin{tikzpicture}
        \begin{groupplot}[group style={group size=5 by 1, y descriptions at=edge left, horizontal sep=5pt}, width=2.75cm, height=\plotHeight, ybar=0pt, ymajorgrids, ylabel=Score, ylabel near ticks, ymin=0, ymax=1, xtick=\empty, legend entries={#2}, legend columns=4, legend to name=lastLegend, nodes near coords, every node near coord/.append style={anchor=east, rotate=90}]
            \nextgroupplot[title=F1]
            \addplot+[discard if not={id}{1}] table[col sep=comma, x expr=0, y=f1] {#1};
            \addplot+[discard if not={id}{2}] table[col sep=comma, x expr=0, y=f1] {#1};
            \addplot+[discard if not={id}{3}] table[col sep=comma, x expr=0, y=f1] {#1};
            \nextgroupplot[title=Recall]
            \addplot+[discard if not={id}{1}] table[col sep=comma, x expr=0, y=rec] {#1};
            \addplot+[discard if not={id}{2}] table[col sep=comma, x expr=0, y=rec] {#1};
            \addplot+[discard if not={id}{3}] table[col sep=comma, x expr=0, y=rec] {#1};
            \nextgroupplot[title=Precision]
            \addplot+[discard if not={id}{1}] table[col sep=comma, x expr=0, y=pre] {#1};
            \addplot+[discard if not={id}{2}] table[col sep=comma, x expr=0, y=pre] {#1};
            \addplot+[discard if not={id}{3}] table[col sep=comma, x expr=0, y=pre] {#1};
            \nextgroupplot[title=MCC]
            \addplot+[discard if not={id}{1}] table[col sep=comma, x expr=0, y=mcc] {#1};
            \addplot+[discard if not={id}{2}] table[col sep=comma, x expr=0, y=mcc] {#1};
            \addplot+[discard if not={id}{3}] table[col sep=comma, x expr=0, y=mcc] {#1};
            \nextgroupplot[title=Accuracy]
            \addplot+[discard if not={id}{1}] table[col sep=comma, x expr=0, y=acc] {#1};
            \addplot+[discard if not={id}{2}] table[col sep=comma, x expr=0, y=acc] {#1};
            \addplot+[discard if not={id}{3}] table[col sep=comma, x expr=0, y=acc] {#1};
        \end{groupplot}
    \end{tikzpicture}}

\def\plotFour#1#2{\begin{tikzpicture}
        \begin{groupplot}[group style={group size=5 by 1, y descriptions at=edge left, horizontal sep=5pt}, width=2.75cm, height=\plotHeight, ybar=0pt, ymajorgrids, ylabel=Score, ylabel near ticks, ymin=0, ymax=1, xtick=\empty, legend entries={#2}, legend columns=4, legend to name=lastLegend, nodes near coords, every node near coord/.append style={anchor=east, rotate=90}]
            \nextgroupplot[title=F1, bar width=8.5pt]
            \addplot+[discard if not={id}{1}] table[col sep=comma, x expr=0, y=f1] {#1};
            \addplot+[discard if not={id}{2}] table[col sep=comma, x expr=0, y=f1] {#1};
            \addplot+[discard if not={id}{3}] table[col sep=comma, x expr=0, y=f1] {#1};
            \addplot+[discard if not={id}{4}] table[col sep=comma, x expr=0, y=f1] {#1};
            \nextgroupplot[title=Recall, bar width=8.5pt]
            \addplot+[discard if not={id}{1}] table[col sep=comma, x expr=0, y=rec] {#1};
            \addplot+[discard if not={id}{2}] table[col sep=comma, x expr=0, y=rec] {#1};
            \addplot+[discard if not={id}{3}] table[col sep=comma, x expr=0, y=rec] {#1};
            \addplot+[discard if not={id}{4}] table[col sep=comma, x expr=0, y=rec] {#1};
            \nextgroupplot[title=Precision, bar width=8.5pt]
            \addplot+[discard if not={id}{1}] table[col sep=comma, x expr=0, y=pre] {#1};
            \addplot+[discard if not={id}{2}] table[col sep=comma, x expr=0, y=pre] {#1};
            \addplot+[discard if not={id}{3}] table[col sep=comma, x expr=0, y=pre] {#1};
            \addplot+[discard if not={id}{4}] table[col sep=comma, x expr=0, y=pre] {#1};
            \nextgroupplot[title=MCC, bar width=8.5pt]
            \addplot+[discard if not={id}{1}] table[col sep=comma, x expr=0, y=mcc] {#1};
            \addplot+[discard if not={id}{2}] table[col sep=comma, x expr=0, y=mcc] {#1};
            \addplot+[discard if not={id}{3}] table[col sep=comma, x expr=0, y=mcc] {#1};
            \addplot+[discard if not={id}{4}] table[col sep=comma, x expr=0, y=mcc] {#1};
            \nextgroupplot[title=Accuracy, bar width=8.5pt]
            \addplot+[discard if not={id}{1}] table[col sep=comma, x expr=0, y=acc] {#1};
            \addplot+[discard if not={id}{2}] table[col sep=comma, x expr=0, y=acc] {#1};
            \addplot+[discard if not={id}{3}] table[col sep=comma, x expr=0, y=acc] {#1};
            \addplot+[discard if not={id}{4}] table[col sep=comma, x expr=0, y=acc] {#1};
        \end{groupplot}
    \end{tikzpicture}}

\section{Introduction}
\label{sec:intro}

% \begin{center}
%     {\itshape The code used for the following experiments is available online at \url{https://github.com/itrummer/DataCorrelationPredictionWithNLP}.}
% \end{center}

Consider a table named ``cars'' with columns named ``maker'' and ``model''. Most people would assume, based on column names \rev{and commonsense knowledge}, that maker and model columns are correlated (i.e., knowing the maker will restrict options for the model). \rev{Such reasoning is possible if column names are meaningful. Assigning meaningful column names is good practice, but of course there are rare exceptions which we are not concerned with here.} In this paper, we study the question of whether automated tuning tools could apply a similar kind of reasoning, exploiting recent innovations in the domain of natural language analysis (NLP): pre-trained language models~\cite{Devlin2019}. 

This research question is motivated by \rev{my recent work}~\cite{Trummer2021nlp, TrummerProfiling}, suggesting to use NLP on database schema elements to inform database tuning, in particular, to help prioritizing data profiling operations. The underlying hypothesis behind those suggestions, namely, whether language models are able to infer relevant information with sufficiently high reliability, has not been investigated in detail. This paper closes that gap, focusing on extracting information about data correlations.

%suggest exploiting natural language processing (NLP) on database schema elements to inform data profiling approaches (``NLP-enhanced data profiling''). 

Detecting correlations in data has been a topic of significant interest in the database research community~\cite{Brown2003, Ilyas2004}. Knowing data correlation is useful in many scenarios. For instance, query optimizers~\cite{Selinger1979} (as well as other tuning tools) often depend on accurate predictions of intermediate result sizes. Classical prediction models assume uncorrelated data, thereby being misled in practice~\cite{Gubichev2015}. As pointed out in prior work~\cite{Ilyas2004}, knowing about correlations can help to correct cardinality estimates. Alternatively, knowing about possible data correlations can help to prune options with correlation-related uncertainty from the search space (e.g., the optimizer can favor join orders where intermediate result sizes do not depend on columns that are likely correlated). 

Detecting data correlations requires comparing data in different columns, often making correlation detection more expensive than operations that focus on different columns in separation. This has motivated dedicated research on algorithms that make correlation detection more efficient~\cite{Brown2003, Ilyas2004}. Typically, those prior algorithms do not exploit information gained via analysis of the database schema, using language models. However, as suggested \rev{in my prior work}~\cite{Trummer2021nlp, TrummerProfiling}, such analysis could be helpful in order to better allocate and prioritize profiling efforts. For instance, given a limited profiling budget, the analysis scope could be restricted to column subsets that are more likely to be correlated, based on the results of NLP. Within those column subsets, any of the existing algorithms for correlation detection could be used. This assumes, however, that NLP is indeed useful to extract relevant information form the database schema. Whether or not that is actually the case, is the subject of the current study.

The hope of extracting useful information from database schema names alone is filled by recent advances in the field of natural language processing. Primarily, those advances are due to two key developments: a novel neural network architecture, the so-called Transformer~\cite{Vaswani2017}, as well as new training methods that exploit large amounts of unlabeled training data~\cite{ruder2019transfer}. Among other advantages, Transformer models enable efficient training of large neural network models with hundreds of millions~\cite{Devlin2019} to hundreds of billions~\cite{Chowdhery2022, Floridi2020} of trainable parameters. Generating task-specific training data at sufficiently large scale is often prohibitively expensive. Fortunately, it is typically possible to reduce the required amount of task-specific training significantly by a pre-training stage that uses large amounts of unlabeled data (e.g., Web text)~\cite{Howard2018}. This study evaluates pre-trained Transformer models, fine-tuned with a moderate amount of training data that is specific to the task of correlation detection. Whereas large Transformer models with hundreds of billions of parameters are nowadays available, typically hosted remotely by providers such as OpenAI~\cite{Floridi2020}, this study focuses on much smaller models (with parameter counts in the hundreds of millions ``only'') that can be run with moderate overheads on commodity machines. This seems reasonable as overheads due to using large language models may otherwise eclipse data profiling overheads altogether.

%Our goal is to automatically identify pairs of columns that are likely correlated, before looking at the actual data and using column names alone. We propose to train deep neural networks for that task. Until quite recently, learning to predict correlations would have required very significant amounts of training data. We propose, however, to exploit an approach that has recently advanced the state of the art in NLP significantly: pre-trained language models~\cite{Devlin2019}. Such models are pre-trained on tasks such as masked language modelling (i.e., predicting masked words based on surrounding text) where large amounts of training data (i.e., any written text) are available. After pre-training, the number of training samples required for specialization to new tasks is very significantly reduced~\cite{Howard2018}. 

This study is based on a newly generated benchmark for data correlation detection. Prior benchmarks of algorithms for correlation detection typically use a small number of data sets~\cite{Ilyas2004}. This is reasonable, as long as performance depends on data properties but not on data semantics. When analyzing column names via language models, however, the data domain may have significant impact on prediction performance (e.g., benefiting application domains that appear more frequently in the pre-training data). Hence, to evaluate language models under realistic conditions, this study uses a benchmark generated from around 4,000 tabular data sets, downloaded from the Kaggle platform. For those data sets, the benchmark analyzes correlation between column pairs according to multiple popular correlation metrics, namely Pearson correlation~\cite{Wang2013g}, Spearman's correlation coefficient~\cite{Artusi2002}, and Theil's U~\cite{theilsu}. While this data is useful to test the primary hypothesis evaluated in this paper, i.e.\ that relevant information can be extracted from schema elements via language models, it can also be used to test NLP-enhanced data profiling approaches. We will see one example of that in Section~\ref{sec:tuning}.

In summary, the original scientific contributions in this experimental paper are the following.

%For our experiments, we use a diverse range of several thousand data sets, considering around 60,000 unique column pairs. We compare different pre-trained language models, as well as simple baselines, and consider different pre

\begin{itemize}
    \item The paper introduces a new benchmark, useful to test correlation prediction, based on column names, and to evaluate approaches for NLP-enhanced database tuning.
    \item The paper tests the ability of language models to infer information on data correlation from column names, considering different correlation metrics, scenarios, and models.
    \item The paper evaluates a simple baseline algorithm for efficient correlation prediction, exploiting information gained via natural language analysis.
\end{itemize}

The remainder of this paper is organized as follows. Section~\ref{sec:related} provides background on the techniques used throughout the paper and discusses related work. Section~\ref{sec:setup} describes the generation of the benchmark, used to evaluate correlation detection methods. Next, Section~\ref{sec:statistics} analyzes the benchmark data set, in terms of data statistics and correlation properties. Section~\ref{sec:methods} compares different methods for predicting data correlations from column names, including pre-trained models and simpler baselines. Section~\ref{sec:scenarios} studies the impact of several scenario properties, including the amount and quality of training data, to study the impact on prediction performance. Section~\ref{sec:breakdown} analyzes prediction performance for different data subsets separately, breaking down, for instance, by column name length among other properties. Section~\ref{sec:metrics} considers different correlation metrics, thereby obtaining insights into how well the prior findings generalize. Finally, Section~\ref{sec:types} evaluates the impact of column types on prediction performance. %An extended technical report is available\footnote{\url{http://itrummer.org/drafts/CorrelationPrediction.pdf}}, reporting additional results in the appendix.}

%we sketch out possible use cases for NLP-based correlation predication and discuss extensions to the evaluated prediction methods. We conclude with a summary of our results and future work directions. In the appendix, we present more detailed results, corroborating our findings. %show results for an extended range of scenarios.
\section{Background and Related Work}
\label{sec:related}

This section discusses prior work, related to this study. Section~\ref{sub:dataprofiling} discusses prior work on data profiling, a primary application domain for the approaches evaluated in this paper. Section~\ref{sub:detectingcorrelation} discusses, more specifically, prior work on data correlation analysis. Section~\ref{sub:languagemodels} discusses the technology that this study is based upon: pre-trained language models. Finally, Section~\ref{sub:nlpfordb} discusses prior work applying such or similar technology in the context of data management.

%We discuss different categories of background and prior work. First, we discuss prior work on database tuning (one of the motivating use cases for this work) and correlation detection. Then, we discuss pre-trained natural language models. Finally, we discuss applications of NLP techniques in the context of database management systems.

\subsection{Data Profiling}
\label{sub:dataprofiling}

The goal of data profiling is to generate statistics and meta-data about a given data set~\cite{Naumann2013}. Specialized tools have been developed for data profiling, including systems from industry~\cite{IBM2021, Informatica2021} as well as academia~\cite{Brown2003, Ilyas2004, Papenbrock2015, Peng2021}. Typically, users specify a target data set for profiling as well as specific types of meta-data to consider. Data profiling is expensive and may have to be repeated periodically as the data changes. Hence, profiling tools often allow users to restrict profiling overheads, e.g.\ by setting time limits~\cite{Papenbrock2015, Talend2021}. %This motivates approaches that prioritize profiling efforts. %, including the one proposed here. 

Profiling methods have been proposed for mining different kinds of meta-data, ranging from statistics over single columns~\cite{Cormode2011a} to more expensive operations such as unique column combination discovery~\cite{Abedjan2014, Papenbrock2017}, detecting inclusion dependencies~\cite{Papenbrock2015a}, foreign keys~\cite{Rostin2009}, order dependencies~\cite{Jin2021, Langer2016}, or statistical data correlations~\cite{Brown2003, Ilyas2004}, the focus of this study.

% \subsection{Database Tuning}

% The performance of database systems is influenced by various tuning choices. Those tuning choices include, for instance, join order, optimized by the query optimizer~\cite{Selinger1979}, as well as physical design decisions. The latter category includes index selection~\cite{Gupta1997, Caprara1995a, Valentin2000}, materialized view selection~\cite{AsgharzadehTalebi2013, Mistry2000}, or partitioning choices~\cite{Nehme2011, Taft2014}, among others. Automated tuning components recommend choices for all of the aforementioned decisions to reduce processing overheads (e.g., for an example workload). Processing cost often depends on the sizes of intermediate results, generated during data processing. Data correlations make it hard to estimate the sizes of those results~\cite{Gubichev2015}. Hence, knowledge on data correlations is valuable to make more reliable tuning choices~\cite{Ilyas2004}. 

\subsection{Detecting Correlations}
\label{sub:detectingcorrelation}

The fact that data correlations are important has motivated work aimed at finding correlations in data sets~\cite{Brown2003, Ilyas2004}. \rev{To guide profiling efforts, such tools typically analyze data samples. The sample size is often chosen as a function of total data size. In contrast, the time for predicting correlation based on column names does not depend on the data size.} Significant work has been dedicated to the problem of selectivity estimation with correlations~\cite{Bruno2004, Markl2007, Tzoumas}. Here, correlations play an important role in estimating aggregate selectivity of predicate groups. More recently, machine learning has been proposed as a method to solve various types of tuning problems in the context of databases~\cite{Marcus2018a, Park2020, Trummer2021c, Wang2021, Woltmann}. Correlated data is a primary reason to replace more traditional cost models, often based on the independence assumption, via learned models. This stream of work connects to this study as it applies machine learning for predicting correlations. However, this study uses machine learning in the form of NLP-based analysis of database schema elements.

\subsection{Language Models}
\label{sub:languagemodels}

Pre-trained language models, based on the Transformer architecture~\cite{Vaswani2017}, have recently led to significant advances on a multitude of NLP tasks~\cite{Wolf2020}. Pre-trained language models are based on the idea of ``transfer learning''. For many specialized NLP tasks, it is difficult to accumulate a large enough body of training data. Also, overheads related to the training of large neural networks from scratch can be significant. This motivates a pre-training step, training a Transformer model on an NLP task for which training data is in ample supply. For instance, this includes the ``masked language modelling'' task~\cite{Devlin2019}. Here, the goal is to predict masked words in a sentence. Doing so requires many capabilities that are useful for other NLP tasks as well. Furthermore, any written text can be used as training data for the latter task. After pre-training, the resulting network (with associated weights) can be specialized (``fine-tuning'') to another task. This refinement step requires only moderate computational resources and few training samples~\cite{Howard2018}, compared to the initial training. Nowadays, pre-trained language models~\cite{Liu2019, Lan2019, Sanh2019} achieve state of the art performance over a wide range of NLP tasks. While the largest models have nowadays hundreds of billions of trainable parameters~\cite{Floridi2020}, this study focuses on smaller models that run on today's commodity machines, making them practical to guide data profiling operations with moderate overheads.

\subsection{NLP for Databases}
\label{sub:nlpfordb}

There have been significant efforts to leverage NLP techniques for database systems~\cite{Trummer2022e}. Typically, the focus of those efforts is the query interface. Here, a long standing goal in database research is to enable natural language query interfaces~\cite{Androutsopoulos1995, Arbor2016, Karagiannis2020a, Karagiannis2020d, Li2014, Sen2020, Saha2016}. Sequence-to-sequence models have been successfully used for this task over the past years~\cite{Hwang2019, Zhong2017, Xu2017}. Recently, pre-trained language models, based on the Transformer architecture~\cite{Vaswani2017}, have achieved excellent results on text-to-SQL benchmarks such as Spider~\cite{Yu2020c} or WikiSQL~\cite{Zhong2017}. They form the basis for this study as well. Other applications of language models in the context of databases include data discovery and integration~\cite{Kayali2023, Narayan2022} as well as data preparation tasks~\cite{Tang2021}.

This study connects to prior work exploiting language models to support the database backend. For instance, this includes work leveraging such models to write code for data processing or process data directly~\cite{Arora2023, Thorne2021, Trummer2022b} or to parse technical documentation to support automated database tuning~\cite{Trummer2022}. More specifically, this study is motivated by \rev{my} prior work suggesting the use of language models on database schema elements to support database tuning and data profiling~\cite{TrummerProfiling, Trummer2021nlp}.

%This work applies NLP not for the interface but rather for the database backend. It thereby connects to another, recently accepted paper~\cite{Trummer2021}. The latter work describes experimental results for a prototype system that parses database tuning hints from text documents. In this work, we present experimental results for predicting data correlations from schema elements instead.

\section{Benchmark}
\label{sec:setup}

This section describes a benchmark for correlation prediction and NLP-enhanced database tuning, created specifically for the purpose of this study. This benchmark covers a wide variety of data sets from different domains. This ensures that the results on prediction performance are representative. 

\subsection{Benchmark Data}
\label{sub:BenchmarkGeneration}

The benchmark uses data sets from the Kaggle Web site\footnote{\url{www.kaggle.com}} (using a corresponding API\footnote{\url{https://github.com/Kaggle/kaggle-api}}). \rev{The choice of Kaggle data is motivated by the large number and diversity of data sets, available on that platform. At the same time, Kaggle data is used for analysis by many data scientists\footnote{\url{https://www.kaggle.com/code/carlmcbrideellis/kaggle-in-numbers}}. Any approach that works well on Kaggle data is likely to benefit a large number of users. As a potential drawback, since Kaggle data is often discussed on the Web, it is possible that data used for pre-training language models contains references to Kaggle data. However, determining data correlation (the prediction target of this benchmark) requires additional data analysis, beyond mere access to data. To the best of my knowledge, no large-scale correlation analysis with similar correlation metrics has been conducted on Kaggle data, prior to the time period during which the models used for experiments were pre-trained. This makes it unlikely that pre-training data contains relevant information on Kaggle data correlation.}

%limit analysis overheads, we consider data sets only up to a size of one megabyte. %For the resulting data sets, we simply download a given number of data sets in the order in which they are returned by the API.

Data sets are obtained by querying the Kaggle API for data sets with the following filters. First, data sets are filtered based on their format, retrieving data sets in ``.csv'' format (i.e., tabular data). Second, to enable retrieval and analysis of a large number of data sets, covering various domains, a size limit of one megabyte was used. The benchmark integrates several thousand data sets, taken from the result of this retrieval query (in the order in which they are returned by the Kaggle API). For those data sets, the benchmark contains various correlation metrics for column pairs within the same table (obtained by analyzing the corresponding data). \rev{Prior work reports that considering correlation between column pairs, as opposed to correlation between more than two columns, ``can remove most of the correlation-induced selectivity estimation errors''~\cite{Ilyas2004}. At the same time, the number of possible correlations grows exponentially in the number of correlated columns, making it more expensive to search for multi-column correlation. Hence, the benchmark focuses on discovering correlation between column pairs.} For each table, only up to 100 column pairs are analyzed. \rev{More precisely, for tables with more than ten columns, the first ten columns are selected for the benchmark (thereby enabling 100 pairs).} The benchmark contains each column pair only once. More precisely, for any given columns $c_1$ and $c_2$, the benchmark contains only one of $\langle c_1,c_2\rangle$ or $\langle c_2,c_1\rangle$ (but not both). While some of the correlation metrics we consider are not symmetric, this avoids using training samples that are too similar (thereby, potentially, leading to overly optimistic prediction performance results). For each column pair, the benchmark contains column names and different correlation metrics. The correlation analysis was executed using Python~3 and SciPy's stats package\footnote{\url{https://docs.scipy.org/doc/scipy/reference/stats.html}}.

The benchmark measures correlation according to different metrics. First, it contains results for the Pearson correlation coefficient~\cite{Wang2013g}. This coefficient is a measure of linear correlation between two data sets. The coefficient itself, denoted as $R$, is contained in the interval $[-1,1]$. It comes with a p-value, indicating the probability of obtaining a specific $R$ value by chance. The following experiments define correlation via different thresholds for $|R|$ while typically using a threshold of 5\% for the p-Value.

Beyond Pearson correlation, measuring linear dependencies, the benchmark also considers Spearman's correlation coefficient~\cite{Hoeffding1957}. This coefficient, typically denoted as $\rho$, measures how well the relationship between two data sets can be characterized by a monotonic (but not necessarily linear) function. Again, the coefficient takes values from the interval $[-1,1]$. It comes with a p-Value, indicating the probability of observing given correlations for uncorrelated data. The following experiments consider different thresholds on $\rho$ while typically requiring a p-value of 5\% or less (to qualify as statistically significant correlation). \rev{}

The two aforementioned coefficients, Pearson's and Spearman's coefficient, apply to numerical columns. For categorical columns (in addition to numerical ones), the benchmark measures the entropy coefficient~\cite{Publications2018}, also called Theil's U, instead. This coefficient is a normalized version of the mutual information between two variables. Intuitively, it measures how many bits we can predict for one column, given the value in the other one. The following experiments vary the threshold on Theil's U, starting from which we consider two columns correlated.

\rev{For all coefficients, the benchmark models correlation prediction as a binary classification problem (classifying column pairs as correlated or uncorrelated, based on the column names). E.g., as an alternative, it is also possible to formulate correlation prediction as a regression problem, aiming to predict correlation metrics such as p-Values or raw coefficient values. Arguably, this is a more challenging problem, as a perfect predictor for the regression variant yields a perfect predictor for the classification version but not vice-versa. At the same time, the classification variant has practical applications, e.g., to select or prioritize column pairs for profiling. Section~\ref{sec:tuning} of the extended technical reports results for a proof-of-concept system, focusing on that use case. For those reasons, this paper focuses on the classification variant and leaves the regression version for future work.}

The result of data preparation is a benchmark, containing the names of columns pairs, meta-data such as the column data type, as well as  correlation results according to different correlation metrics.

\subsection{Benchmark Metrics}
\label{sub:benchmarkmetrics}

The following experiments use the benchmark data for two types of experiments. First, the experiments evaluate the ability of language models to predict data correlation from column names. Second, the experiments evaluate a simple algorithm for NLP-enhanced data profiling, exploiting predictions on data correlation to prioritize data profiling steps.

To measure the ability of language models to predict data correlation, the experiments measure prediction quality according to multiple metrics. More precisely, the experiments consider five metrics of prediction quality: recall, precision, and the F1 score (which combines recall and precision). Here, recall is the percentage of correlated column pairs that were accurately identified. Precision is the ratio of actual column pairs among the ones predicted to be correlated. The F1 score is defined as $2\cdot p\cdot r/(p+r)$ (where $p$ and $r$ are precision and recall, respectively). The aforementioned three metrics are typically used in scenarios where a relatively sparse class of elements should be identified. Strongly correlated column pairs qualify as they tend to be relatively sparse (as shown in the next section).

The experiments also measure Matthew's Correlation Coefficient (MCC)~\cite{Chicco2020} and simple prediction accuracy (considering the two classes ``correlated'' and ``uncorrelated'' for each column pair). All of the aforementioned quality metrics yield values from the interval $[0,1]$ and higher values represent better quality. The following plots report all five metrics. When verifying hypotheses about prediction quality, we consider a hypotheses as validated, if it is validated according to all of those five metrics.

The performance of NLP-enhanced data profiling tools is evaluated by the number of correlated column pairs, verified within a budget on computation overheads. This budget can be measured, for instance, by the number of column pairs analyzed (the benchmark contains time measurements for correlation analysis as well, enabling experiments with budgets on computation time). Section~\ref{sec:tuning} contains more details on this scenario.
\section{Benchmark Analysis}
\label{sec:statistics}

\begin{table*}
    \centering
    \caption{Examples for strongly correlated (top) and less correlated (bottom) columns according to Pearson's coefficient.}
    \label{tab:examples}
    \begin{tabular}{lllll}
        \toprule[1pt]
        \textbf{Data Set} & \textbf{Column 1} & \textbf{Column 2} & \textbf{$R$ value} & \textbf{$p$ value} \\
        \midrule[1pt]
         epl1920leaguetable.csv &  Points & Wins & 99\% & 0\% \\
         emission data.csv &  1751 & 1752 & 100\% & 0\% \\
         india-districts-census-2011.csv & Female\_Literate & Male\_Literate & 98\% & 0\% \\
         Google\_Stock\_Price\_Test.csv & High & Open & 96\% & 0\% \\
         housing.csv & total\_bedrooms & total\_rooms & 93\% & 0\% \\
         \midrule[1pt]
         time\_series\_2019-ncov-Confirmed.csv &  1/24/20 0:00 & Lat & -6\% & 83\% \\
         diabetes\_merged\_date-time-sorted-includes-patient-id.csv & code & patient\_id & 2\% & 0\% \\
         Heart.csv & RestECG & Sex & 2\% & 71\% \\
         2020.12.09/2020.12.09.csv & num\_pkts\_out & dest\_ip & -5\% & 0\% \\
         nyc-east-river-bicycle-counts.csv & Williamsburg Bridge & Unnamed: 0 & 10\% & 16\% \\
         \bottomrule[1pt]
    \end{tabular}
\end{table*}

This section analyzes the benchmark, introduced in the previous section. Table~\ref{tab:examples} shows an extract from this data set. The upper part of the table shows five highly correlated columns, measured via the Pearson coefficient, the lower half shows five column pairs with low correlation. At the same time, it shows column names and the names of the associated data sets. 

For the examples in the table, it seems often possible, using commonsense knowledge, to identify likely candidates for correlations. For instance, the number of points for a team in a table with sports statistics often correlates with the number of wins (more points often lead to more wins). Indeed, the corresponding column pair shows relatively high correlation that is statistically significant (using the common thresholds of 5\% to separate statistically significant from non-significant p-Values). On the other side, there is no obvious indication of any correlation between two columns named ``Williamsburg Bridge'' and ``Unnamed: 0''. Indeed, the corresponding column pair shows only weak correlation. This paper studies the question whether language models are able to simulate such reasoning.

\begin{table}[t]
    \caption{Statistics on benchmark data sets.}
    \label{tab:statistics}
    \centering
    \begin{tabular}{ll}
    \toprule[1pt]
    \textbf{Property} & \textbf{Value} \\
    \midrule[1pt]
    Number of data sets & 3,952 \\
    \midrule
    Number of column pairs & 119,384 \\
    \midrule
    Number of numerical column pairs & 59,449 \\
    \midrule
    Number of rows (Avg.) & 103,126 \\
    \midrule
    Number of distinct values per column (Avg.) & 6,200 \\
    \bottomrule[1pt]
    \end{tabular}
\end{table}

Table~\ref{tab:statistics} summarizes size-related statistics, describing the benchmark data. Altogether, the benchmark contains correlations from about 4,000 data sets. Those data sets derive from various sources and cover various topics (the examples in Table~\ref{tab:examples} give a first impression of their diversity). For those data sets, the benchmark contains results about 120K column pairs, about half of them of numerical (or integer) type. In average, the source data sets contain over 100K rows and each column contains over 6K unique values.

\begin{figure}
    \centering
    \begin{tikzpicture}
        \begin{groupplot}[group style={group size=2 by 1, y descriptions at=edge left, horizontal sep=10pt}, ylabel={Count}, width=4.5cm, height=\plotHeight, ybar=0pt, ymajorgrids, ymode=log, ymax=250000, ymin=1, xmin=0]
            \nextgroupplot[bar width=6pt, xlabel=$\log_2(Chars)$]
            \addplot table[x index=0, y index=1] {results/analysis/lengthh};
            \nextgroupplot[bar width=8.5pt, xlabel=$\log_2(Tokens)$]
            \addplot table[x index=0, y index=1] {results/analysis/tokensh};
        \end{groupplot}
    \end{tikzpicture}
    \caption{Distribution of column name length, measured as the number of characters (left) and number of tokens (right).}
    \label{fig:nameAnalysis}
\end{figure}

The ability to predict likely correlation from column names may depend on features such as the column name length. Intuitively, having longer column names should be more informative. Potentially, this makes correlation prediction easier. Figure~\ref{fig:nameAnalysis} shows histograms summarizing the distribution of column name length, measured according to different metrics. The left plot shows the distribution over character length. The right plot measures the number of tokens (i.e., text snippets separated by spaces or underscores) in column pairs. Note the logarithmic x-axis. The average column name length is around 16 characters. This is sufficient for few, short words. The number of tokens is typically limited to two (i.e., one word in each of the columns). %Later, in Section~\ref{sec:breakdown}, we analyze how column name properties correlate with prediction accuracy.

\begin{figure}
    \centering
    \begin{tikzpicture}
        \begin{groupplot}[group style={group size=3 by 1, y descriptions at=edge left, horizontal sep=10pt}, ylabel={Count}, width=3.5cm, height=\plotHeight, ybar=0pt, ymajorgrids, ymode=log, ymax=100000, ymin=1, xmin=0, xtick={0, 5, 10}, xticklabels={0,0.5,1}, xmin=-0.5, xmax=10.5]
            \nextgroupplot[bar width=5pt, xlabel=Pearson ($|R|$)]
            \addplot table[x index=0, y index=1] {results/analysis/pearsonh};
            \nextgroupplot[bar width=5pt, xlabel=Spearman ($|\rho|$)]
            \addplot table[x index=0, y index=1] {results/analysis/spearmanh};
            \nextgroupplot[bar width=5pt, xlabel={Theil's U}]
            \addplot table[x index=0, y index=1] {results/analysis/theilsuh};
        \end{groupplot}
    \end{tikzpicture}
    \caption{Distribution of correlation coefficient values according to different metrics.}
    \label{fig:correlationAnalysis}
\end{figure}

The following experiments vary the threshold, starting from which columns are considered correlated. This makes it interesting to analyze how correlation is distributed over column pairs. Figure~\ref{fig:correlationAnalysis} shows histograms, characterizing the corresponding distributions. From left to right, it shows correlation according to Pearson's coefficient, Spearman's coefficient, and according to Theil's U. In particular for Pearson's and Spearman's coefficients, low values are more likely than higher ones. For all correlation metrics, we see a slightly bimodal distribution with increased probability for maxima and minima.
\section{Comparing Prediction Methods}
\label{sec:methods}

This section compares different prediction methods in terms of their training time (if any) and output quality. Most of them use pre-trained language models, based on the recently proposed Transformer architecture~\cite{Vaswani2017}. Such models achieve state of the art performance in a variety of NLP tasks~\cite{Howard2018}. %Next, we describe all compared prediction methods. Afterwards, we present corresponding results.

\subsection{Description of Methods}

The experiments consider three pre-trained models, small enough to be used locally on today's commodity machines. \rev{All of them are encoder models, pre-trained to associate input tokens with high-dimensional vectors. For classification problems, a thin layer is added that maps vectors to scores for the relevant classes.} Roberta~\cite{Liu2019} (short for ``Robustly Optimized BERT approach'') expands BERT~\cite{Devlin2019}, another pre-trained language model that has achieved widespread popularity. Compared to BERT, Roberta is pre-trained using more data and for a longer period of time. The resulting model outperforms the original BERT model on various benchmarks. 

Albert (short for ``A lite BERT architecture'')~\cite{Lan2019} reduces the number of parameters, compared to BERT and Roberta, significantly. It uses two parameter reduction techniques. First, it decomposes the vocabulary embedding matrix into two smaller matrices. Second, it shares parameters across different layers (thereby reducing parameter growth as a function of network depth). This model achieves significant speedups without affecting result quality significantly.

Distilbert~\cite{Sanh2019}, a ``distilled'' version of BERT, uses knowledge distillation to reduce parameters and training time. As suggested by the name, it uses knowledge distillation to reduce the network size, compared to BERT. Here, BERT serves as a ``teacher'' that trains a smaller network, Distilbert. The authors of Distilbert show \rev{that the resulting model realizes attractive tradeoffs between training time and result quality.}

Beyond pre-trained models, the evaluation considers a simple baseline. This baseline decomposes names of compared columns into tokens. Then, it computes the Jaccard similarity on the two token sets, associated with the compared columns. It predicts a correlation if the Jaccard similarity is at least 0.5. Hence, this baseline considers columns as correlated if their names are sufficiently similar.

Note that the evaluation focuses uniquely on methods that work with column names alone, as opposed to methods that exploit the actual data for correlation analysis. As demonstrated in Section~\ref{sec:tuning}, methods of the former category can be used to guide application of methods that belong to the latter category.

\subsection{Experimental Setup}
\label{sub:HardwareSoftwareSetup}

The experiments in this and the following sections use an EC2 instance of type p3.2xlarge, recommended for machine learning workloads. It features a Tesla V100 GPU with 5,120 CUDA cores, 8~vCPUs, and 488~GB of RAM. Prediction methods are implemented using the simpletransformers Python library\footnote{\url{https://simpletransformers.ai/}}, using the default parameter settings of that library, unless noted otherwise. The simpletransformers library is internally based on the Huggingface library\footnote{\url{https://huggingface.co/transformers/}} which supports a wide range of pre-trained language models. %The following experiments compare different models, using the Roberta model~\cite{Liu2019} (``roberta-base'') as default, as well as . As alternatives, we also consider Albert~\cite{Lan2019} (``albert-base-v2'') and DistilBERT~\cite{Sanh2019} (``distilbert-base-cased''), two more efficient alternatives. By default, we train for five epochs (a setting that works well for most NLP tasks in the context of pre-trained models~\cite{Liu2019}), using training and test batches of size 100. We use the GPU for training.

\subsection{Comparison Results}

%This subsection reports experimental results, comparing different methods, to evaluate the following hypotheses.

The following experiments use the Pearson correlation coefficient. Two columns are considered correlated if $|R|\geq 0.9$ with a p-value of at most 5\%. For training, 80\% of numerical column pairs are used (around 47K pairs) while reporting results for the remaining 20\%. Prediction quality is measured according to all metrics introduced in Section~\ref{sub:benchmarkmetrics}. %Also, we measure training time (for the baselines with training phase).

\def\plotFourChanged#1#2{\begin{tikzpicture}
        \begin{groupplot}[group style={group size=5 by 1, y descriptions at=edge left, horizontal sep=5pt}, width=2.75cm, height=\plotHeight, ybar=0pt, ymajorgrids, ylabel=Score, ylabel near ticks, ymin=0, ymax=1, xtick=\empty, legend entries={#2}, legend columns=4, legend to name=lastLegend, nodes near coords, every node near coord/.append style={anchor=east, rotate=90}]
            \nextgroupplot[title=F1, bar width=8.5pt]
            \addplot+[discard if not={id}{1}, every node near coord/.append style={anchor=west}] table[col sep=comma, x expr=0, y=f1] {#1};
            \addplot+[discard if not={id}{2}] table[col sep=comma, x expr=0, y=f1] {#1};
            \addplot+[discard if not={id}{3}] table[col sep=comma, x expr=0, y=f1] {#1};
            \addplot+[discard if not={id}{4}] table[col sep=comma, x expr=0, y=f1] {#1};
            \nextgroupplot[title=Recall, bar width=8.5pt]
            \addplot+[discard if not={id}{1}, every node near coord/.append style={anchor=west}] table[col sep=comma, x expr=0, y=rec] {#1};
            \addplot+[discard if not={id}{2}] table[col sep=comma, x expr=0, y=rec] {#1};
            \addplot+[discard if not={id}{3}] table[col sep=comma, x expr=0, y=rec] {#1};
            \addplot+[discard if not={id}{4}] table[col sep=comma, x expr=0, y=rec] {#1};
            \nextgroupplot[title=Precision, bar width=8.5pt]
            \addplot+[discard if not={id}{1}] table[col sep=comma, x expr=0, y=pre] {#1};
            \addplot+[discard if not={id}{2}] table[col sep=comma, x expr=0, y=pre] {#1};
            \addplot+[discard if not={id}{3}] table[col sep=comma, x expr=0, y=pre] {#1};
            \addplot+[discard if not={id}{4}] table[col sep=comma, x expr=0, y=pre] {#1};
            \nextgroupplot[title=MCC, bar width=8.5pt]
            \addplot+[discard if not={id}{1}, every node near coord/.append style={anchor=west}] table[col sep=comma, x expr=0, y=mcc] {#1};
            \addplot+[discard if not={id}{2}] table[col sep=comma, x expr=0, y=mcc] {#1};
            \addplot+[discard if not={id}{3}] table[col sep=comma, x expr=0, y=mcc] {#1};
            \addplot+[discard if not={id}{4}] table[col sep=comma, x expr=0, y=mcc] {#1};
            \nextgroupplot[title=Accuracy, bar width=8.5pt]
            \addplot+[discard if not={id}{1}] table[col sep=comma, x expr=0, y=acc] {#1};
            \addplot+[discard if not={id}{2}] table[col sep=comma, x expr=0, y=acc] {#1};
            \addplot+[discard if not={id}{3}] table[col sep=comma, x expr=0, y=acc] {#1};
            \addplot+[discard if not={id}{4}] table[col sep=comma, x expr=0, y=acc] {#1};
        \end{groupplot}
    \end{tikzpicture}}

\begin{figure}
    \centering
    \plotFourChanged{results/performance/methodscmp.csv}{Baseline, Albert, Distilbert, Roberta}
    
    \ref{lastLegend}
    \caption{Comparison of correlation prediction methods.}
    \label{fig:methods}
\end{figure}

\begin{hypothesis}\label{hyp:bettermodels}
Pre-trained language models predict correlations better than simpler baselines.
\end{hypothesis}

Figure~\ref{fig:methods} compares prediction quality across the four prediction methods. The simple baseline performs quite well (even though not as good as the other methods) for precision and accuracy. Here, the baseline benefits as it predicts no correlation in most cases. As correlated column pairs are rare, this simple strategy can achieve a relatively high accuracy (which, among other things, motivates the use of multiple quality metrics). The baseline predicts a correlation for columns with very similar names. It seems that such column pairs tend to be correlated indeed, explaining the reasonably high precision values. However, the simple baseline achieves only poor results for recall, F1 score, and the MCC metric. For instance, its recall is around 1\% only. This demonstrates the need for a more sophisticated approach (thereby validating Hypothesis~\ref{hyp:bettermodels}). 

\begin{hypothesis}\label{hyp:largerquality}
Larger models predict correlations more reliably than smaller models.
\end{hypothesis}

The performance of all three pre-trained models is quite similar. For the F1 score, precision, MCC, and recall, Roberta performs best, even though only by a small margin. This is expected as Roberta is the largest of the three models. For recall, Distilbert has a slight advantage. In general, Distilbert performs slightly better than Albert in the experiments. MCC is the metric with the largest gap between the three models. Here, Roberta gains a performance advantage of 4\% over Albert. Altogether, the three models realize however comparable prediction performance (providing only weak evidence for Hypothesis~\ref{hyp:largerquality}).

\begin{figure}
    \centering
    \begin{tikzpicture}
        \begin{axis}[legend entries={Distilbert, Albert, Roberta}, legend pos=outer north east, width=5cm, height=\plotHeight, ybar=0pt, ymin=0, ymajorgrids, xtick=\empty, ylabel={Time (minutes)}, nodes near coords, every node near coord/.append style={anchor=east, rotate=90}, bar width=10pt]
            \addplot coordinates {(0,13)};
            \addplot coordinates {(0,17)};
            \addplot coordinates {(0,25)};
        \end{axis}
    \end{tikzpicture}
    \caption{Training time of different transformer variants.}
    \label{fig:times}
\end{figure}

\begin{hypothesis}\label{hyp:largertime}
Varying the model size enables different tradeoffs between computational overheads and accuracy.
\end{hypothesis}

Figure~\ref{fig:times} reports training time (in minutes) for the three pre-trained models. More precisely, it reports the time for fine-tuning the three models to the problem of correlation prediction (i.e., it does not report time for pre-training which is significantly more expensive). 

Albert and Distilbert have been designed with the goal of reducing overheads, compared to larger models such as Roberta. The results indicate that this approach pays off for the correlation prediction task as well. For instance, training time is more than two times smaller for Distilbert, compared to Roberta. Given the slightly better performance of Distilbert, compared to Albert, this model seems like the best alternative to reduce training overheads. In the following, due to slightly higher precision, Roberta is used as default method for correlation prediction. \rev{Overall, the results support Hypothesis~\ref{hyp:largertime}.}

\section{Scenario Variants}
\label{sec:scenarios}

The following experiments analyze how scenario properties influence prediction performance. \rev{This section} considers variations in the amount of training data as well as in the relationship between training and test data. Also, it compares performance for different definitions of correlation. %More precisely, the following experiments evaluate the following hypotheses.

\begin{figure}
    \centering

    \begin{subfigure}[b]{\columnwidth}
        \centering
        \plotThree{results/performance/tquantcmp.csv}{80\% Training, 20\% Training, No Training}
        
        \ref{lastLegend}
        \caption{\rev{Impact of training data quantity.}}
        \label{fig:tquantity}        
    \end{subfigure}

    \begin{subfigure}[b]{\columnwidth}
        \centering
        \plotTwo{results/performance/tqualicmp.csv}{Separate by Data Set, Separate by Column Pair}
        
        \ref{lastLegend}
        \caption{Impact of training data quality.}
        \label{fig:tquality}
    \end{subfigure}

    \begin{subfigure}[b]{\columnwidth}
        \centering
        \plotFour{results/performance/strongcmp.csv}{$|R|\geq0.8$, $|R|\geq0.9$, $|R|\geq0.95$, $|R|\geq0.99$}
        
        \ref{lastLegend}
        \caption{Impact of correlation strength.}
        \label{fig:strength}
    \end{subfigure}
    
    \caption{\rev{Prediction quality for Pearson correlation coefficient in different scenarios.}}
    \label{fig:pearsonScenarios}
\end{figure}

The following experiments use the Roberta model and define correlation via the Pearson coefficient, using a maximal p-Value of 5\%. The threshold on the absolute value of $R$ varies in the following, using a default of $|R|\geq0.9$. The quantity of training data varies as well, using 80\% of numerical column pairs as default. The remaining data is used for testing.

% \begin{figure}
%     \centering
%     \plotTwo{results/performance/tquantcmp.csv}{80\% Training Data, 20\% Training Data}
    
%     \ref{lastLegend}
%     \caption{Impact of training data quantity.}
%     \label{fig:tquantity}
% \end{figure}

\begin{hypothesis}\label{hyp:quantity}
Correlation predictions become more accurate when training on more column pairs.
\end{hypothesis}

Figure~\ref{fig:tquantity} reports results related to Hypothesis~\ref{hyp:quantity}. It compares prediction performance as a function of the training (and test) ratio. It compares performance in two scenarios. The first scenario uses 80\% of column pairs as training data (i.e., around 47K column pairs) and the rest for testing. The second scenario uses 20\% of column pairs as training data while using the rest for testing. \rev{The third scenario uses no training data whatsoever (i.e., zero-shot setting).} For all five metrics of prediction quality, as expected, having more training data helps. This validates Hypothesis~\ref{hyp:quantity}.

However, given the significant difference in the amount of training data (the amount of training data differs by a factor of four across the two scenarios), the differences in prediction performance seem moderate. The maximal difference across all five performance metrics is six percent. This is consistent with prior results for other tasks from the NLP domain, showing that pre-trained language models achieve reasonable performance, already with modest amounts of training data~\cite{Howard2018}. 

% \begin{figure}
%     \centering
%     \plotTwo{results/performance/tqualicmp.csv}{Separate by Data Set, Separate by Column Pair}
    
%     \ref{lastLegend}
%     \caption{Impact of training data quality.}
%     \label{fig:tquality}
% \end{figure}

\begin{hypothesis}\label{hyp:quality}
Predicting correlations for new column pairs becomes easier after observing correlations from other column pairs in the same data set.
\end{hypothesis}

The following analysis focus on relationship between training and test data on prediction performance. It considers two scenarios. The first scenario separates training and test data at the granularity of column pairs. This means that training and test data may contain column pairs from the same data set. Of course, each column pair is considered only once. The second scenario ensures that training and test data are derived from different data sets. This means separating data sets (the ones used for correlation analysis) into training and test data sets, then deriving column pairs for training and testing only from the corresponding data sets. With this method, training and test samples derive from entirely different data sets. \rev{Of course, training and test data sets may still have similarities. For instance, the same column names may appear in different data sets. However, even if column names are identical, the associated data (and therefore the correlation results) may still differ. Overall, since similarities between data sets are common beyond Kaggle, the results for the second scenario are of high practical relevance. They illustrate the performance obtained by training predictors on a representative collection of data sets, then applying them to new data sets (while benefiting from naturally occurring similarities between training data and new data).}

Figure~\ref{fig:tquality} shows corresponding results. It varies the quality (i.e., how closely it relates to the test examples) but not the quantity of training data. For all five quality metrics, allowing column pairs from the same data set in training and testing improves performance. Depending on the metric, this increase is moderate (recall increase by only 2\%) or more significant (precision increases by 14\%). In any case, the results support Hypothesis~\ref{hyp:quality}. 

% \begin{figure}
%     \centering
%     \plotFour{results/performance/strongcmp.csv}{$|R|\geq0.8$, $|R|\geq0.9$, $|R|\geq0.95$, $|R|\geq0.99$}
    
%     \ref{lastLegend}
%     \caption{Comparing prediction quality for different degrees of correlation.}
%     \label{fig:strength}
% \end{figure}

\begin{hypothesis}\label{hyp:strength}
Predicting strong correlations is easier than predicting weak correlations.
\end{hypothesis}

Hypothesis~\ref{hyp:strength} connects the criterion for defining correlation to prediction performance. The following experiment uses a p-Value threshold of 5\% but varies the threshold on $|R|$ between $0.8$ and $0.99$. Figure~\ref{fig:strength} shows corresponding results. The results do not show clear tendencies. Recall generally increases while precision and F1 scores decrease, as the requirements for correlation become tighter. Accuracy and MCC do not show clear tendencies. Altogether, the experimental results do not provide strong evidence for Hypothesis~\ref{hyp:strength}.
\section{Result Breakdowns}
\label{sec:breakdown}

This section explores the question of which properties of test cases contribute to making correlations more or less difficult to predict, breaking down results based on properties of column names. %More specifically, it focuses on the following two hypotheses.

\begin{figure}
    \centering

    \begin{subfigure}[b]{\columnwidth}
        \centering
        \plotThree{results/performance/nrchars.csv}{Q0-25\%, Q25-50\%, Q50-100\%}
        
        \ref{lastLegend}
        \caption{Impact of the number of characters.}
        \label{fig:nrchars}
    \end{subfigure}

    \begin{subfigure}[b]{\columnwidth}
        \centering
        \plotThree{results/performance/nrtokens.csv}{Q0-25\%, Q25-50\%, Q50-100\%}
        
        \ref{lastLegend}
        \caption{Impact of the number of words.}
        \label{fig:nrwords}
    \end{subfigure}

    \begin{subfigure}[b]{\columnwidth}
        \centering
        \plotThree{results/performance/wordratio.csv}{Q0-25\%, Q25-50\%, Q50-100\%}
        
        \ref{lastLegend}
        \caption{Impact of ratio of English words in column names.}
        \label{fig:wordratio}
    \end{subfigure}
    
    \caption{\rev{Breakdown of prediction quality by test case properties for the Pearson correlation coefficient.}}
    \label{fig:lengthImpact}
\end{figure}

% \begin{figure}
%     \centering
%     \plotThree{results/performance/nrchars.csv}{Q0-25\%, Q25-50\%, Q50-100\%}
    
%     \ref{lastLegend}
%     \caption{Impact of column name length.}
%     \label{fig:nrchars}
% \end{figure}

% \begin{figure}
%     \centering
%     \plotThree{results/performance/nrtokens.csv}{Q0-25\%, Q25-50\%, Q50-100\%}
    
%     \ref{lastLegend}
%     \caption{Impact of number of words in column names.}
%     \label{fig:nrwords}
% \end{figure}

The following experiments use the Roberta model and the Pearson correlation coefficient. They consider columns correlated for an absolute R-value of at least $0.9$ and a p-Value of at most 5\%. 

\begin{hypothesis}
Longer column names yield more information and make prediction more accurate.\label{hyp:namelength}
\end{hypothesis}

The following experiments focus on Hypothesis~\ref{hyp:namelength} and consider two metrics of column name length: the number of characters and the number of tokens (i.e., the number of text snippets, separated by spaces and underscores). Figure~\ref{fig:nrchars} reports results for the number of characters and Figure~\ref{fig:nrwords} results for the number of tokens. Both figures report results for three length ranges separately. Those ranges refer to the quantiles (e.g., the range ``Q50-100\%'' includes pairs of columns whose length is at or above average length).

The differences are significant. For both column name metrics and all five prediction quality metrics, having longer column names improves performance. Those differences are more pronounced when measuring length as the number of characters (compared to the number of words). For instance, when measuring length as the number of characters, the MCC score improves from 42\% to 83\% when going from short to long names. Precision improves from 45\% to 81\%. Those results provide strong experimental evidence validating Hypothesis~\ref{hyp:namelength}. \rev{Shorter column names may indicate a higher ratio of ``placeholder'' names (e.g., based on column numbers) or correlate with less detailed explanations of column semantics. The next hypothesis relates  to the nature of column names.}

% \begin{figure}
%     \centering
%     \plotThree{results/performance/wordratio.csv}{Q0-25\%, Q25-50\%, Q50-100\%}
    
%     \ref{lastLegend}
%     \caption{Impact of ratio of English words in column names.}
%     \label{fig:wordratio}
% \end{figure}

\begin{hypothesis}
Column names with a higher ratio of English words (as opposed to abbreviations or other symbols) can be more easily interpreted and make predictions more accurate.\label{hyp:wordratio}
\end{hypothesis}

Figure~\ref{fig:wordratio} reports results for subsets of column pairs, characterized by the ratio of English words in column names. It measures that ratio as follows. First, both column names are divided into tokens, using common separators. Then, each token is compared to an English dictionary. The ratio of words, contained in the dictionary, to the number of all tokens is the word ratio. For Figure~\ref{hyp:wordratio} separates column pairs into three groups, associated with different ranges for that ratio (e.g., ``Q50-100\%'' includes column pairs whose ratio of English words is at or above the average). 

Here, the absolute differences between low and high word ratios are relatively small. For instance, MCC scores increase only from 60\% to 65\%, moving from low to high ratios. Recall, for instance, decreases slightly from 88\% to 87\%. Altogether, the experimental evidence for Hypothesis~\ref{hyp:wordratio} is weak.

\rev{The results in this section also show that it is possible to assess the confidence of correlation predictions, based on properties of column names (specifically: the length). This may be useful for systems exploiting correlation prediction as a component, as discussed in more detail in Section~\ref{sec:tuning} of the extended technical report.}
\section{Other Correlation Metrics}
\label{sec:metrics}

This section expands the experimental scope from Pearson correlation to other correlation metrics, thereby verifying whether prior findings generalize. The following experiments consider Spearman's coefficient and Theil's U (discussed in more detail in Section~\ref{sec:setup}). Unless noted otherwise, they consider columns correlated, according to Spearman's coefficient, if the absolute coefficient value is at or above 0.9 ($|\rho|\geq0.9$) with a p-value of at most 5\%. For Theil's U, it uses a threshold of 0.9 as well. By default, it uses again 80\% of column pairs as training data and separate training from test data at the granularity of column pairs. Note that Theil's U applies to all column data types, thereby increasing the number of eligible column pairs (for training and testing) to around 119,000. 

\begin{figure}
    \centering
    \plotThree{results/performance/metricscmp.csv}{Pearson, Spearman, Theil's U}
    
    \ref{lastLegend}
    \caption{Comparison of different correlation measures.}
    \label{fig:metrics}
\end{figure}

Figure~\ref{fig:metrics} compares prediction performance for all three definitions of correlation. While prediction performance is close for Pearson and Spearman coefficients, prediction performance increases for Theil's U according to all quality metrics. A first hypothesis is that the higher amount of training data (columns of all types as opposed to numerical columns only) contributes to that performance. %It hints at the possibility to improve prediction performance for the other coefficients further by increasing the amount of training data (i.e., more numerical column pairs) as well.

\begin{figure}
    \centering
    \begin{subfigure}[b]{\columnwidth}
        \centering
        \plotThree{results/performance/sptquantcmp.csv}{80\% Training, 20\% Training, No Training}
        
        \ref{lastLegend}
        \caption{\rev{Impact of training data quantity.}}
        \label{fig:sptquantity}
    \end{subfigure}

    \begin{subfigure}[b]{\columnwidth}
        \centering
        \plotTwo{results/performance/sptqualicmp.csv}{Separate by Data Set, Separate by Column Pair}
        
        \ref{lastLegend}
        \caption{Impact of training data quality.}
        \label{fig:sptquality}
    \end{subfigure}

    \begin{subfigure}[b]{\columnwidth}
        \centering
        \plotFour{results/performance/spstrongcmp.csv}{$|\rho|\geq0.8$, $|\rho|\geq0.9$, $|\rho|\geq0.95$, $|\rho|\geq 0.99$}
    
        \ref{lastLegend}
        \caption{Impact of degree of correlation.}
        \label{fig:spstrength}
    \end{subfigure}
    
    \caption{\rev{Prediction quality for Spearman's coefficient in different prediction scenarios.}}
    \label{fig:spearmanScenarios}
\end{figure}

% \begin{figure}
%     \centering
%     \plotTwo{results/performance/sptquantcmp.csv}{80\% Training Data, 20\% Training Data}
    
%     \ref{lastLegend}
%     \caption{Impact of training data quantity when predicting correlation according to Spearman's coefficient.}
%     \label{fig:sptquantity}
% \end{figure}

% \begin{figure}
%     \centering
%     \plotTwo{results/performance/sptqualicmp.csv}{Separate by Data Set, Separate by Column Pair}
    
%     \ref{lastLegend}
%     \caption{Impact of training data quality when predicting correlation according to Spearman's coefficient.}
%     \label{fig:sptquality}
% \end{figure}

% \begin{figure}
%     \centering
%     \plotFour{results/performance/spstrongcmp.csv}{$|\rho|\geq0.8$, $|\rho|\geq0.9$, $|\rho|\geq0.95$, $|\rho|\geq 0.99$}
    
%     \ref{lastLegend}
%     \caption{Comparing prediction quality for different degrees of correlation, measured via Spearman's coefficient.}
%     \label{fig:spstrength}
% \end{figure}

\rev{Figure~\ref{fig:spearmanScenarios} re-tests} the hypotheses from Section~\ref{sec:scenarios} for Spearman's coefficient. Clearly, increasing the amount of training data also increases prediction performance (Hypothesis~\ref{hyp:quantity}), as well as sharing data sets among training and test cases (Hypothesis~\ref{hyp:quality}). The tendencies are less clear for the threshold on $\rho$ (Hypothesis~\ref{hyp:strength}).

\begin{figure}
    \centering

    \begin{subfigure}[b]{\columnwidth}
        \centering
        \plotThree{results/performance/spnrchars.csv}{Q0-25\%, Q25-50\%, Q50-100\%}
        
        \ref{lastLegend}
        \caption{Impact of column name length.}
        \label{fig:spnrchars}
    \end{subfigure}

    \begin{subfigure}[b]{\columnwidth}
        \centering
        \plotThree{results/performance/spnrtokens.csv}{Q0-25\%, Q25-50\%, Q50-100\%}
        
        \ref{lastLegend}
        \caption{Impact of number of words in column names.}
        \label{fig:spnrwords}
    \end{subfigure}

    \begin{subfigure}[b]{\columnwidth}
        \centering
        \plotThree{results/performance/spwordratio.csv}{Q0-25\%, Q25-50\%, Q50-100\%}
        
        \ref{lastLegend}
        \caption{Impact of ratio of English words in column names.}
        \label{fig:spwordratio}
    \end{subfigure}
    
    \caption{\rev{Breakdown of prediction quality by test case properties for Spearman's coefficient.}}
    \label{fig:spearmanBreakdowns}
\end{figure}

% \begin{figure}
%     \centering
%     \plotThree{results/performance/spnrchars.csv}{Q0-25\%, Q25-50\%, Q50-100\%}
    
%     \ref{lastLegend}
%     \caption{Impact of column name length when predicting correlation according to Spearman's coefficient.}
%     \label{fig:spnrchars}
% \end{figure}

% \begin{figure}
%     \centering
%     \plotThree{results/performance/spnrtokens.csv}{Q0-25\%, Q25-50\%, Q50-100\%}
    
%     \ref{lastLegend}
%     \caption{Impact of number of words in column names when predicting correlation according to Spearman's coefficient.}
%     \label{fig:spnrwords}
% \end{figure}

% \begin{figure}
%     \centering
%     \plotThree{results/performance/spwordratio.csv}{Q0-25\%, Q25-50\%, Q50-100\%}
    
%     \ref{lastLegend}
%     \caption{Impact of ratio of English words in column names when predicting correlation according to Spearman's coefficient.}
%     \label{fig:spwordratio}
% \end{figure}

\rev{Figure~\ref{fig:spearmanBreakdowns} validates} the hypotheses from Section~\ref{sec:breakdown} for Spearman's coefficient. It considers different data subsets and compare prediction performance. Again, the most important parameter influencing prediction performance seems to be the column name length (Hypothesis~\ref{hyp:namelength}). 

%%% Theil's U

\begin{figure}
    \centering

    \begin{subfigure}[b]{\columnwidth}
        \centering
        \plotThree{results/performance/tutquantcmp.csv}{80\% Training, 20\% Training, No Training}
        
        \ref{lastLegend}
        \caption{\rev{Impact of training data quantity.}}
        \label{fig:tutquantity}
    \end{subfigure}

    \begin{subfigure}[b]{\columnwidth}
        \centering
        \plotTwo{results/performance/tutqualicmp.csv}{Separate by Data Set, Separate by Column Pair}
        
        \ref{lastLegend}
        \caption{Impact of training data quality.}
        \label{fig:tutquality}
    \end{subfigure}

    \begin{subfigure}[b]{\columnwidth}
        \centering
        \plotTwo{results/performance/tustrongcmp.csv}{$U\geq0.8$, $U\geq0.9$}
        
        \ref{lastLegend}
        \caption{Impact of degree of correlation.}
        \label{fig:tustrength}
    \end{subfigure}
    
    \caption{\rev{Prediction quality for Theil's U in different prediction scenarios.}}
    \label{fig:theilsUscenarios}
\end{figure}

% \begin{figure}
%     \centering
%     \plotTwo{results/performance/tutquantcmp.csv}{80\% Training Data, 20\% Training Data}
    
%     \ref{lastLegend}
%     \caption{Impact of training data quantity when predicting correlation according to Theil's U.}
%     \label{fig:tutquantity}
% \end{figure}

% \begin{figure}
%     \centering
%     \plotTwo{results/performance/tutqualicmp.csv}{Separate by Data Set, Separate by Column Pair}
    
%     \ref{lastLegend}
%     \caption{Impact of training data quality when predicting correlation according to Theil's U.}
%     \label{fig:tutquality}
% \end{figure}

% \begin{figure}
%     \centering
%     \plotTwo{results/performance/tustrongcmp.csv}{$U\geq0.8$, $U\geq0.9$}
    
%     \ref{lastLegend}
%     \caption{Comparing prediction quality for different degrees of correlation, measured via Theil's U.}
%     \label{fig:tustrength}
% \end{figure}

\rev{Figure~\ref{fig:theilsUscenarios} compares} different prediction scenarios for Theil's U. Here, the relative tendencies are similar to prior experiments while the absolute values are significantly better. It is interesting that prediction quality for Theil's U, when using 20\% training data (i.e., around 24K training samples), is still better than prediction performance for the other coefficients when using 80\% of training data (i.e., around 47K training samples). This shows that, beyond the amount of training data, other factors must contribute to the improved performance. %We plan to analyze those factors in future work. 

\begin{figure}
    \centering

    \begin{subfigure}[b]{\columnwidth}
        \centering
        \plotThree{results/performance/tunrchars.csv}{Q0-25\%, Q25-50\%, Q50-100\%}
        
        \ref{lastLegend}
        \caption{Impact of column name length.}
        \label{fig:tunrchars}
    \end{subfigure}

    \begin{subfigure}[b]{\columnwidth}
        \centering
        \plotThree{results/performance/tunrtokens.csv}{Q0-25\%, Q25-50\%, Q50-100\%}
        
        \ref{lastLegend}
        \caption{Impact of number of words in column names.}
        \label{fig:tunrwords}
    \end{subfigure}

    \begin{subfigure}[b]{\columnwidth}
        \centering
        \plotThree{results/performance/tuwordratio.csv}{Q0-25\%, Q25-50\%, Q50-100\%}
        
        \ref{lastLegend}
        \caption{Impact of ratio of English words in column names.}
        \label{fig:tuwordratio}
    \end{subfigure}
    
    \caption{\rev{Breakdown of prediction quality by test case properties for Theil's U.}}
    \label{fig:theilsUbreakdowns}
\end{figure}

% \begin{figure}
%     \centering
%     \plotThree{results/performance/tunrchars.csv}{Q0-25\%, Q25-50\%, Q50-100\%}
    
%     \ref{lastLegend}
%     \caption{Impact of column name length when predicting correlation according to Theil's U.}
%     \label{fig:tunrchars}
% \end{figure}

% \begin{figure}
%     \centering
%     \plotThree{results/performance/tunrtokens.csv}{Q0-25\%, Q25-50\%, Q50-100\%}
    
%     \ref{lastLegend}
%     \caption{Impact of number of words in column names when predicting correlation according to Theil's U.}
%     \label{fig:tunrwords}
% \end{figure}

% \begin{figure}
%     \centering
%     \plotThree{results/performance/tuwordratio.csv}{Q0-25\%, Q25-50\%, Q50-100\%}
    
%     \ref{lastLegend}
%     \caption{Impact of ratio of English words in column names when predicting correlation according to Theil's U.}
%     \label{fig:tuwordratio}
% \end{figure}

Figures~\ref{fig:tunrchars} to \ref{fig:tuwordratio} study prediction performance for Theil's U and different data subsets. While column name length remains the most important factor, a higher ratio of English words in column names relates to better prediction accuracy as well (except for the precision metric). Despite of that, the absolute differences remain relatively small. 

Altogether, the primary outcomes of prior experiments generalize to other definitions of correlation. %This provides evidence for the potential of NLP-based correlation prediction methods in a variety of scenarios.

\section{Column Types}
\label{sec:types}

\rev{This section focuses on column types and their impact on prediction quality.}

\begin{hypothesis}\label{hyp:typeimpact}
    \rev{Prediction quality varies as a function of column types.}
\end{hypothesis}

\rev{Tables~\ref{tab:pearsonpertype} to \ref{tab:theilsupertype} break down prediction results according to the data types of the involved columns. Column types are inferred automatically by the pandas framework, considering types bool (B), float64 (F), int64 (I), and object (O). The first column of each table contains the types of column pairs (e.g., ``I-F'' indicates a column pair where the first column is of type int64 whereas the second column is of type float64). Table~\ref{tab:pearsonpertype} breaks down results for Pearson correlation, Table~\ref{tab:spearmanpertype} reports results for the Spearman correlation coefficient, and Table~\ref{tab:theilsupertype} reports results for Theil's U.}

\begin{table}
    \centering
    \caption{\rev{Impact of types on Pearson correlation prediction.}}
    \label{tab:pearsonpertype}
    \begin{tabular}{llllll}
    \toprule[1pt]
    Types & F1 & Pre & Rec & Acc & MCC \\
    \midrule[1pt]
F-F & 0.88 & 0.84 & 0.93 & 0.93 & 0.84 \\
F-I & 0.67 & 0.58 & 0.78 & 0.98 & 0.66 \\
I-F & 0.85 & 0.83 & 0.88 & 0.99 & 0.84 \\
I-I & 0.62 & 0.49 & 0.85 & 0.81 & 0.54 \\
\bottomrule[1pt]
    \end{tabular}
\end{table}

\begin{table}
    \centering
    \caption{\rev{Impact of types on Spearman correlation prediction.}}
    \label{tab:spearmanpertype}
    \begin{tabular}{llllll}
    \toprule[1pt]
    Types & F1 & Pre & Rec & Acc & MCC \\
    \midrule[1pt]
F-F & 0.87 & 0.82 & 0.93 & 0.93 & 0.83 \\
F-I & 0.68 & 0.60 & 0.78 & 0.97 & 0.67 \\
I-F & 0.71 & 0.63 & 0.81 & 0.97 & 0.70 \\
I-I & 0.56 & 0.41 & 0.90 & 0.76 & 0.49 \\
\bottomrule[1pt]
    \end{tabular}
\end{table}

\begin{table}
    \centering
    \caption{\rev{Impact of types on Theil's U correlation prediction.}}
    \label{tab:theilsupertype}
    \begin{tabular}{llllll}
    \toprule[1pt]
    Types & F1 & Pre & Rec & Acc & MCC \\
    \midrule[1pt]
O-O & 0.92 & 0.92 & 0.92 & 0.90 & 0.80 \\
O-F & 0.93 & 0.92 & 0.93 & 0.90 & 0.79 \\
O-I & 0.94 & 0.95 & 0.93 & 0.92 & 0.83 \\
O-B & 0.50 & 0.33 & 1.00 & 0.86 & 0.53 \\
\midrule
F-O & 0.95 & 0.95 & 0.94 & 0.93 & 0.85 \\
F-F & 0.95 & 0.96 & 0.95 & 0.93 & 0.80 \\
F-I & 0.94 & 0.94 & 0.94 & 0.93 & 0.86 \\
F-B & 0.00 & 0.00 & 0.00 & 0.80 & 0.00 \\
\midrule
I-O & 0.93 & 0.92 & 0.93 & 0.92 & 0.85 \\
I-F & 0.94 & 0.96 & 0.93 & 0.93 & 0.85 \\
I-I & 0.80 & 0.89 & 0.73 & 0.85 & 0.69 \\
I-B & 0.00 & 0.00 & 0.00 & 1.00 & 0.00 \\
\midrule
B-O & 0.89 & 0.84 & 0.93 & 0.85 & 0.69 \\
B-F & 0.80 & 0.67 & 1.00 & 0.80 & 0.67 \\
B-I & 0.96 & 1.00 & 0.92 & 0.95 & 0.90 \\
B-B & 0.18 & 0.11 & 0.50 & 0.67 & 0.10 \\
\bottomrule[1pt]
    \end{tabular}
\end{table}

\rev{Clearly, column types have significant impact on prediction accuracy. E.g., for Pearson correlation, MCC scores vary by 30\% across different column type combinations. The order of types matters (e.g., comparing F-I versus I-F in Table~\ref{tab:pearsonpertype}). This can be explained by the fact that the column position correlates with the likelihood of correlations. For instance, key columns tend to appear first in tables and are often of type integer. Hence, the probability of correlation for an integer column, followed by a float column, is different than for the complementary order. This can influence accuracy and other performance metrics for prediction. At the same time, it motivates extensions of the approach that exploit the column position in addition to column names.}

\rev{Overall, the results clearly support Hypothesis~\ref{hyp:typeimpact} and could be exploited, e.g., to assess the confidence in a prediction, based on column types. At the same time, they raise the question whether column types may be useful as additional input for the language model itself.}

\begin{figure}[t]
    \centering

    \begin{subfigure}[b]{\columnwidth}
        \centering
        \plotTwo{results/typed/typedpearson.csv}{Untyped, Typed}
        
        \ref{lastLegend}
        \caption{Predicting Pearson correlation.}
        \label{fig:pearsontyped}
    \end{subfigure}

    \begin{subfigure}[b]{\columnwidth}
        \centering
        \plotTwo{results/typed/typedspearman.csv}{Untyped, Typed}
        
        \ref{lastLegend}
        \caption{Predicting Spearman correlation.}
        \label{fig:spearmantyped}
    \end{subfigure}

    \begin{subfigure}[b]{\columnwidth}
        \centering
        \plotTwo{results/typed/typedtheilsu.csv}{Untyped, Typed}
        
        \ref{lastLegend}
        \caption{Predicting Theil's U correlation.}
        \label{fig:theilsutyped}
    \end{subfigure}
    
    \caption{\rev{Exploiting types of columns for classification (in addition to column names).}}
    \label{fig:typed}
\end{figure}

\begin{hypothesis}\label{hyp:typefeatures}
    \rev{Integrating column types as additional feature increases prediction accuracy.}
\end{hypothesis}

\rev{Prior experiments have focused on predicting data correlation from column names alone. The following experiment considers column types as an additional feature. Figure~\ref{fig:typed} reports corresponding results. Different from before, the input to the language model now contains column names, followed by column types (separated by a single space). E.g., for two columns ``car'' and ``maker'' of types ``object'', the input consists of the pair ``car object'' and ``maker object''.}

\rev{Figure~\ref{fig:typed} compares results with and without types (using 80\% of data for training). Considering types leads to moderate benefits for most correlation metrics and in most scenarios. E.g., using types leads to improvements of four percentage points in precision when predicting correlation according to Theil's U. On the other hand, it leads to slight losses when predicting correlation according to Spearman's correlation coefficient. Overall, the results provide weak support for Hypothesis~\ref{hyp:typefeatures}.}

% Interestingly, MCC scores are uniformly better when using typed inputs. Also, precision increases significantly in most cases. On the other hand, recall suffers. This means the model exploits data types to avoid predicting correlations for column pairs that seem likely candidates for correlation, based on column names alone. Both of the aforementioned effects nearly cancel each other out, resulting in comparable F1 scores. Accuracy decreases when adding types. Hence, whether adding types is preferable depends on the requirements of the specific use case.}

%\input{sections/implications}
\section{Conclusion}
\label{sec:conclusion}

\rev{In recent publications, I suggest} using advanced natural language analysis on text associated with database schema elements. This is a cheap source of information as the cost depends only on the schema, but not on the data size. Ideally, natural language analysis yields insights on likely data properties, that are helpful to guide automated tuning or data profiling efforts.

\rev{This suggestion is based} on the assumption that pre-trained language models are indeed able to extract useful insights from schema text. For the first time, this study evaluates that hypothesis in detail, focusing on the problem of correlation detection. Correlation detection is an expensive process that has received significant attention in the database community, due to its various use cases in database optimization. Hence, obtaining additional information to guide corresponding profiling efforts is practically useful.

%\rev{This paper explores the use of NLP to predict data properties, namely correlation, from column names.} 

The experiments yields the following insights (among others):

\begin{itemize}
    \item In many, even though not all, cases, pre-trained language models are able to infer useful information on data correlation from column names alone.
    \item This is already possible with relatively small models, e.g.\ \rev{Distilbert}, with parameter counts in the tens of millions, enabling their use on commodity machines.
    \item Those findings hold for a variety of popular data correlation metrics, including Pearson correlation, Spearman correlation, and Theil's U.
    \item Training models for correlation prediction on data sets that are similar to test data increases performance, motivating domain specialization.
    \item Surprisingly, prediction accuracy is only marginally affected by the degree of data correlation.
    \item On the other hand, predictions become more accurate if more text is available, i.e.\ if column names are longer.
    % \item Using NLP-based predictions enables data profiling tools to prioritize profiling efforts, thereby allowing them to find more results than baselines, given the same profiling budget.
\end{itemize}

The experimental results inform future research aimed at NLP-enhanced database tuning and data profiling. They provide evidence supporting assumptions underlying that nascent research direction.

\bibliographystyle{ACM-Reference-Format}
\bibliography{library}

\newpage
\clearpage

\appendix
\section{NLP-Enhanced Data Profiling}
\label{sec:tuning}

Language models are able, in principle, to extract information about likely data correlations from column names. This section tries to leverage that capability to make data profiling more efficient. Section~\ref{sub:context} describes a model for NLP-enhanced data profiling, based on prior publications on that topic~\cite{Trummer2021nlp, TrummerProfiling}. Section~\ref{sub:baselines} presents \rev{several simple profiling baselines} that exploit NLP to optimize allocation of profiling steps. Finally, Section~\ref{sub:results} presents results for \rev{those baselines}, based on the benchmark data described in Section~\ref{sec:setup}.

\subsection{Context}
\label{sub:context}

\tikzstyle{component}=[draw, rectangle, shade, top color=blue!10, bottom color=blue!20, minimum width=4cm, font=\bfseries, blur shadow={shadow blur steps=5}]
\tikzstyle{dataflow}=[thick, -latex]
\tikzstyle{io}=[font=\bfseries]

\begin{figure}
    \centering
    \begin{tikzpicture}
        \node[io] (data) at (-3,1) {Data};
        \node[io] (schema) at (0,1) {Database Schema};
        \node[component] (analyze) at (0,0) {Analyze Schema via NLP};
        \node[component] (schedule) at (0,-1) {Schedule Profiling Steps};
        \node[component] (execute) at (0,-2) {Execute Profiling Steps};

        \draw[dataflow] (schema) -- (analyze);
        \draw[dataflow] (analyze) -- (schedule);
        \draw[dataflow] (schedule) -- (execute);
        \draw[dataflow] (data.south) -- (-3,-1) -- (schedule.west);
        \draw[dataflow] (data.south) -- (-3,-2) -- (execute.west);
        \draw[dataflow] (execute.east) -- ++ (1,0) -- ++ (0,1) -- (schedule.east);
    \end{tikzpicture}
    \caption{Architecture for NLP-enhanced data profiling.}
    \label{fig:profilingoverview}
\end{figure}

Figure~\ref{fig:profilingoverview} shows a high-level architecture for NLP-enhanced data profiling, consistent with suggestions from prior work~\cite{Trummer2021nlp, TrummerProfiling}. According to that architecture, NLP can be used to prioritize profiling operations, thereby making data profiling more efficient. NLP is applied to the database schema, analyzing names of columns and tables (possibly, additional text derived, for instance, from comments in the schema definition could be used). This analysis yields insights on likely data properties, e.g., likely data correlations.

Data profiling can be expensive. Hence, typically, data profiling algorithms allow users to specify constraints on resource consumption (e.g., measured as profiling time)~\cite{Naumann2013, Papenbrock2015}. Under such resource constraints, profiling is potentially unable to analyze the entire data set. Then, to maximize the number of profiling results (e.g., to maximize the number of correlated column pairs found~\cite{Brown2003, Ilyas2004}), it is crucial to prioritize profiling steps. This is where natural language analysis of schema element names can help.

In Figure~\ref{fig:profilingoverview}, names in the database schema are analyzed via NLP. The results of this analysis inform a scheduler, planning profiling steps to maximize a utility function (e.g., the number of correlated column pairs uncovered), given resource constraints. The execution component takes care of executing scheduled profiling steps (e.g., a correlation analysis between two specific columns). The results of analysis may, in turn, inform the profiling scheduler when planning the next profiling steps.

%The goal of data profiling~\cite{Naumann2013, Papenbrock2015} is to derive useful meta-data about a data set, including statistics on value distributions in certain columns~\cite{Naumann2013} as well as information on correlations between columns~\cite{Brown2003, Ilyas2004}.

% \subsection{Benchmark}

\subsection{\rev{Baselines}}
\label{sub:baselines}

The following experiments evaluate baseline \rev{algorithms} that instantiate (parts of) the architecture in Figure~\ref{fig:profilingoverview}. The goal is to find correlated column pairs, \rev{according to one of the correlation metrics discussed previously,} given a budget in terms of the number of column pairs to analyze. Note that \rev{those algorithms are} relatively simple. \rev{Their} primary purpose is to establish first baseline results for the problem of NLP-enhanced database tuning on the benchmark presented in this paper.

%, described in a prior publication~\cite{TrummerProfiling} but without detailed evaluation, 

%The first baseline algorithm, ``Basic'' in the following plots, simply analyzes column pairs in random order. It does not yet exploit the results of NLP on schema elements, nor does it exploit any other techniques that make finding correlated columns easier (e.g., utilizing information on column types or column samples~\cite{Brown2003, Ilyas2004}). 

The \rev{first} baseline algorithm \rev{(``LM'' in the following plots)} uses NLP in a simplistic manner to prioritize profiling operations. More precisely, it prioritizes column pairs, predicted to be correlated according to the Roberta model (fine-tuned as explained in the previous sections). Among columns pairs in the same category (i.e., either predicted to be correlated or not correlated), it prioritizes column pairs based on the confidence score of the model prediction. For column pairs predicted to be correlated, it prioritizes column pairs with a higher confidence. For column pairs predicted not to be correlated, it prioritizes pairs with a lower confidence. \rev{Note that this baseline is relatively simple and could be extended in multiple ways. For instance, the results in Section~\ref{sec:breakdown} motivate a mechanism that calculates classification confidence as a function of column name length (and other column properties). This would enable a system that prioritizes column pairs for profiling that combine a high classification confidence with a character length above average. Such refinements are, however, beyond the scope of the current publication.} \rev{The second baseline (``Jaccard'' in the following plots) is a variant of the first that uses Jaccard similarity between column names (considering each column name as a set of tokens) as a proxy for the likelihood of correlation (see Section~\ref{sec:methods} for further experiments with the Jaccard distance). This means that column pairs with high Jaccard similarity are analyzed first.}

\subsection{Results}
\label{sub:results}

\begin{figure}
    \centering
    \begin{tikzpicture}
        \begin{groupplot}[group style={group size=2 by 3, x descriptions at=edge bottom, y descriptions at=edge left}, width=4cm, height=\plotHeight, ymajorgrids, xlabel={Column Pairs}, ylabel={\# Detections}, ymin=0, ymax=0.6, xtick=data, xticklabels={0.05,0.1,0.15,0.2,0.25}, legend entries={LM, Jaccard, Expected}, legend columns=3, legend to name=prioritizedTuningLegend]
        \nextgroupplot[title={Pearson, All Tables}]
            \addplot table[col sep=comma] {results/tuning/pearsonAllLM.txt};
            \addplot table[col sep=comma] {results/tuning/pearsonAllJaccard.txt};
            %\addplot coordinates {(0.05,0.05) (0.1,0.1) (0.15,0.15) (0.2,0.2) (0.25,0.25)};
        \nextgroupplot[title={Pearson, Large Tables}]
            \addplot table[col sep=comma] {results/tuning/pearsonLargeLM.txt};
            \addplot table[col sep=comma] {results/tuning/pearsonLargeJaccard.txt};
            %\addplot coordinates {(0.05,0.05) (0.1,0.1) (0.15,0.15) (0.2,0.2) (0.25,0.25)};
        \nextgroupplot[title={Spearman, All Tables}]
            \addplot table[col sep=comma] {results/tuning/spearmanAllLM.txt};
            \addplot table[col sep=comma] {results/tuning/spearmanAllJaccard.txt};
            %\addplot coordinates {(0.05,0.05) (0.1,0.1) (0.15,0.15) (0.2,0.2) (0.25,0.25)};
        \nextgroupplot[title={Spearman, Large Tables}]
            \addplot table[col sep=comma] {results/tuning/spearmanLargeLM.txt};
            \addplot table[col sep=comma] {results/tuning/spearmanLargeJaccard.txt};
            %\addplot coordinates {(0.05,0.05) (0.1,0.1) (0.15,0.15) (0.2,0.2) (0.25,0.25)};
        \nextgroupplot[title={Theil's U, All Tables}]
            \addplot table[col sep=comma] {results/tuning/theilsuAllLM.txt};
            \addplot table[col sep=comma] {results/tuning/theilsuAllJaccard.txt};
            %\addplot coordinates {(0.05,0.05) (0.1,0.1) (0.15,0.15) (0.2,0.2) (0.25,0.25)};
        \nextgroupplot[title={Theil's U, Large Tables}]
            \addplot table[col sep=comma] {results/tuning/theilsuLargeLM.txt};
            \addplot table[col sep=comma] {results/tuning/theilsuLargeJaccard.txt};
            %\addplot coordinates {(0.05,0.05) (0.1,0.1) (0.15,0.15) (0.2,0.2) (0.25,0.25)};

        \end{groupplot}
    \end{tikzpicture}

    \ref{prioritizedTuningLegend}
    \caption{\rev{Comparing baselines when profiling for correlation according to different metrics.}}
    \label{fig:tuning}
\end{figure}

\rev{Figure~\ref{fig:tuning} evaluates} the aforementioned baselines for correlation detection. \rev{The algorithms are} evaluated for different profiling budgets, shown on the x-axis. The profiling budget limits the number of analyzed column pairs, relative to the total number of column pairs for a specific table. The \rev{LLM algorithm} uses a Roberta model, trained on 80\% of tables (selected with uniform random distribution). The results, reported in the figure, refer to the remaining 20\% of tables. \rev{The Jaccard baseline uses the Jaccard similarity between column names.} On the y-axis, the figure reports the ratio of correlated column pairs that were detected up to a certain profiling budget. In the plots, ``All Tables'' refers to the results for all tables in the test set. ``Large Tables'' refers to the average ratio of detected correlations, considering only tables with at least ten columns. Finally, ``Expected'' refers to the expected ratio of detected correlated column pairs when considering column pairs in random order (it matches the ratio of analyzed column pairs).

Clearly, exploiting the results of NLP enables the algorithm to detect significantly more correlations than expected, given the same profiling budget. \rev{Using Jaccard similarity for prioritization fails for large tables with many columns. Among all data sets that can benefit from prioritization, i.e., data sets with at least one correlated and one uncorrelated column pair, 88\% have at least ten columns (averaging over all correlation metrics). Here, precise predictions obtained via language model pay off, resulting in significant gains.} 

\rev{Predicting correlation via language models adds overheads. On the benchmark platform described in Section~\ref{sub:HardwareSoftwareSetup}, in average, correlation prediction via Roberta takes 3.5 milliseconds per column pair. On the other hand, in average, data analysis takes 165 milliseconds per column pair. Hence, the overheads due to prediction are negligible, compared to the overheads of data analysis. Sub-sampling a small number of rows and analyzing the correlation for that sample is another way to obtain information on correlation with small overheads. However, the sample size relates to the strength of the correlation results\footnote{\url{https://www.cfholbert.com/blog/sample-size-correlation/}}, meaning that larger samples are required for stronger results (increasing analysis overheads). Correlation results obtained on a sub-sample are not necessarily representative. On the other hand, the overheads for NLP neither depend on the number of rows nor on the target correlation metric. A more sophisticated approach may use NLP as well as analysis of data samples as two complementary heuristics to guide profiling actions. This is however beyond the scope of this paper. Altogether, the results provide first evidence that NLP can be a useful source of information in data profiling.}

%\rev{Note that this benchmark focuses on relatively small tables with a size of up to one megabyte and around 100K rows in average (see Sections~\ref{sub:BenchmarkGeneration} and \ref{sec:statistics}). For tables with more rows, the overheads of data analysis per column pair will increase, unlike the overheads for prediction. Altogether, while the evaluated algorithms are simple, the results provide first evidence for the potential of NLP-enhanced data profiling.}

% Less than 8\% of column pairs have a non-zero Jaccard similarity
% 
\section{Prediction Confidence versus Correlation Strength}

\rev{In this section, we analyze dependencies between prediction confidence and correlation strength. We consider the same predictors, the same data, and the same settings as in Section~\ref{sec:tuning}. Section~\ref{sec:tuning} uses Jaccard similarity between column names and the confidence scores of the Roberta model predictions to guide data profiling efforts. The following analysis compares those two metrics, referring to both, Jaccard similarity and model prediction confidence, as ``confidence scores''.}

% In the following plots and tables, ``Jaccard'' refers to the Jaccard similarity between column names while ``LM'' refers to the predictions of the language model (Roberta). 

%In Section~\ref{sec:tuning}, we used two methods to predict data correlation: the Jaccard similarity between column names and the predictions of a language model (Roberta). In this section, we drill down further and analyze 

\begin{table}[t]
    % Associated SQL query: with target as (select jaccard as t, labels, method from predictions) select method, (avg(case when labels= 1 then t else NULL end) - avg(case when labels=0 then t else NULL end))/(select avg(t) from target) from target group by method order by method;
    \centering
    \caption{\rev{Relative increase in prediction confidence, comparing correlated to uncorrelated columns.}}
    \begin{tabular}{lll}
        \toprule[1pt]
        \textbf{Coefficient} & \textbf{$\Delta$ Jaccard} (\%) & \textbf{$\Delta$ LM} (\%) \\
        \midrule[1pt]
         Pearson & 316 & $7\cdot 10^{-4}$ \\
         Spearman & 340 & $762$ \\
         Theil's U & -71 & $5.5\cdot 10^{-2}$ \\
         \bottomrule[1pt]
    \end{tabular}    
    \label{tab:predictionDeltas}
\end{table}

\rev{Table~\ref{tab:predictionDeltas} compares predictions for correlated and non-correlated columns, considering the same three correlation metrics (Pearson, Spearman, and Theil's U) as before. It reports the relative distance in prediction confidence. More precisely, denoting by $c_C$, $c_U$, $c_A$ the average confidence scores for correlated, uncorrelated, and all columns, the table reports the value for the formula $(c_C-c_U)/c_A$ (as a percentage). Ideally, prediction confidence is higher for correlated, compared to uncorrelated columns. This means, for an ideal metric, the values reported in Table~\ref{tab:predictionDeltas} are all positive. This is the case for the predictions by the language model (``LM'' in Table~\ref{tab:predictionDeltas}). For Jaccard similarity (``Jaccard''), it holds for the first two correlation metrics (Pearson and Spearman) but not for the last one (Theil's U). This provides further evidence that confidence scores by the language model are more reliable predictors of correlation, compared to Jaccard similarity (consistent with the results of Sections~\ref{sec:methods} and \ref{sec:tuning}). Note that the relative distance varies significantly for the language model, comparing different correlation metrics. This is explained by the fact that the model is trained separately for predicting correlation according to the three correlation metrics. The three resulting model versions differ significantly by their range of confidence scores. This does however not correlate with their prediction performance.}

%The results in Table~\ref{}

\begin{figure}
    \centering
    \begin{tikzpicture}
        \begin{groupplot}[group style={group size=1 by 3, x descriptions at=edge bottom}, width=8cm, height=\plotHeight, ymajorgrids, xlabel={Prediction Confidence (Percentile)}, ylabel={Mean Coefficient}, legend entries={LM Confidence,Jaccard Similarity}, legend to name=predictionVsCorrelationLegend, legend columns=2]
            \nextgroupplot[title={Pearson Correlation}]
            \addplot table[x index=0, y index=1, col sep=comma] {results/prediction/pearsonperc.csv};
            \addplot[mark=x, draw=red] table[x index=0, y index=3, col sep=comma] {results/prediction/jaccardpearsonperc};
            \nextgroupplot[title={Spearman Correlation}]
            \addplot table[x index=0, y index=1, col sep=comma] {results/prediction/spearmanperc.csv};
            \addplot table[x index=0, y index=3, col sep=comma] {results/prediction/jaccardspearmanperc};
            \nextgroupplot[title={Theil's U Correlation}]
            \addplot table[x index=0, y index=1, col sep=comma] {results/prediction/theilsuperc.csv};
            \addplot table[x index=0, y index=3, col sep=comma] {results/prediction/jaccardtheilsuperc};
        \end{groupplot}
    \end{tikzpicture}

    \ref{predictionVsCorrelationLegend}
    \caption{\rev{Prediction confidence versus correlation strength for different correlation metrics and prediction methods.}}
    \label{fig:predictionVsCorrelation}
\end{figure}

\rev{Figure~\ref{fig:predictionVsCorrelation} provides more details. The x-axis represents percentiles of the confidence scores for correlation. More precisely, the percentiles refer to the scores calculated for ranking purposes, as described in Section~\ref{sub:baselines}. The y-axis represents the average correlation coefficient value (for the three correlation metrics used in the experiments throughout the paper) for all column pairs with prediction confidence above the value, represented on the x-axis. The Jaccard similarity is zero for more than 92\% of column pairs (i.e., column names do not share any tokens). Therefore, Jaccard similarity does not correlate with correlation coefficients for a large part of test cases. On the other hand, coefficient values increase gradually as a function of prediction confidence for the language model, in particular for Pearson and Spearman correlation.}

\section{Results for All Column Pairs}

The main part of the paper presents an extract of the experimental results. This appendix provides additional results, corroborating prior findings. 

The following tables present detailed results for the Pearson correlation coefficient and for all numerical column pairs. All of them define correlation by a p-value of at most 5\% to consider a correlation statistically significant. The experiments reported in the following vary the prediction method (using the three pre-trained models described in Section~\ref{sec:methods}). Also, they vary the definition of correlation by setting different thresholds for $|R|$, ranging from 0.8 to 0.99. Furthermore, the experiments vary the amount of training data used. The following three tables (Tables~\ref{tab:robdef} to \ref{tab:disdef}) report results when separating training from test data at the granularity of single column pairs. 

In this and the following tables, the first column ($|R|$) reports the minimal threshold on the Pearson correlation coefficient $|R|$. The second column denotes the ratio of data used for testing, as opposed to training (i.e., ``0.2'' means that 20\% of data were used for testing while the other 80\% were used for training). The following columns denote the prediction quality metrics introduced in Section~\ref{sec:setup} (i.e., the F1 score, recall, precision, accuracy, and MCC, in that order). Each of the following tables reports results for one of the three pre-trained models.

\begin{table}
    \centering
    \caption{Correlation prediction quality for column pair separation and the Roberta transformer ($p\leq 0.05$).}
    \label{tab:robdef}
    \begin{tabular}{llllllll}
    \toprule[1pt]
    $\mathbf{|R|}$ & Test & F1 & Rec & Pre & Acc & MCC \\
    \midrule[1pt]
0.8 & 0.2 & 72 & 86 & 62 & 86 & 64 \\ \midrule
0.8 & 0.8 & 68 & 84 & 57 & 83 & 59 \\ \midrule
0.9 & 0.2 & 71 & 87 & 61 & 87 & 65 \\ \midrule
0.9 & 0.8 & 67 & 84 & 56 & 85 & 60 \\ \midrule
0.95 & 0.2 & 68 & 91 & 54 & 86 & 63 \\ \midrule
0.95 & 0.8 & 65 & 87 & 52 & 85 & 59 \\ \midrule
0.99 & 0.2 & 65 & 93 & 50 & 87 & 62 \\ \midrule
0.99 & 0.8 & 62 & 89 & 48 & 86 & 58 \\ \bottomrule[1pt]
    \end{tabular}
\end{table}

\begin{table}
    \centering
    \caption{Correlation prediction quality for column pair separation and the Albert transformer ($p\leq 0.05$).}
    \label{tab:albdef}
    \begin{tabular}{llllllll}
    \toprule[1pt]
    $\mathbf{|R|}$ & Test & F1 & Rec & Pre & Acc & MCC \\
    \midrule[1pt]
0.8 & 0.2 & 70 & 81 & 61 & 85 & 61 \\ \midrule
0.8 & 0.8 & 67 & 76 & 60 & 84 & 57 \\ \midrule
0.9 & 0.2 & 68 & 86 & 57 & 85 & 61 \\ \midrule
0.9 & 0.8 & 58 & 66 & 52 & 82 & 48 \\ \midrule
0.95 & 0.2 & 67 & 88 & 54 & 86 & 61 \\ \midrule
0.95 & 0.8 & 66 & 78 & 57 & 87 & 59 \\ \midrule
0.99 & 0.2 & 64 & 91 & 50 & 87 & 61 \\ \midrule
0.99 & 0.8 & 62 & 87 & 49 & 86 & 58 \\
\bottomrule[1pt]
    \end{tabular}
\end{table}

\begin{table}
    \centering
    \caption{Correlation prediction quality for column pair separation and the Distilbert transformer ($p\leq 0.05$).}
    \label{tab:disdef}
    \begin{tabular}{llllllll}
    \toprule[1pt]
    $\mathbf{|R|}$ & Test & F1 & Rec & Pre & Acc & MCC \\
    \midrule[1pt]
0.8 & 0.2 & 72 & 86 & 62 & 85 & 64 \\ \midrule
0.8 & 0.8 & 69 & 81 & 61 & 84 & 60 \\ \midrule
0.9 & 0.2 & 71 & 87 & 60 & 87 & 65 \\ \midrule
0.9 & 0.8 & 68 & 83 & 57 & 85 & 60 \\ \midrule
0.95 & 0.2 & 68 & 90 & 55 & 86 & 63 \\ \midrule
0.95 & 0.8 & 65 & 87 & 51 & 84 & 58 \\ \midrule
0.99 & 0.2 & 65 & 93 & 50 & 87 & 62 \\ \midrule
0.99 & 0.8 & 62 & 88 & 48 & 86 & 58 \\
\bottomrule[1pt]
    \end{tabular}
\end{table}

The results are consistent with the hypotheses established in the main part of the paper. As a general rule, using more training data increases accuracy (Hypothesis~\ref{hyp:quantity}) by a few percentage points for most metrics. Again, tendencies are less clear for the threshold on the Pearson coefficient. The F1 scores typically decrease once the definition of correlation becomes more restrictive. This is mostly driven by a decrease in precision while recall even increases in several cases. On the other side, accuracy remains quite stable and even increases in some cases, as the number of correlated columns shrinks. 

Comparing the three models, Roberta turns out to be the most accurate one in most cases. The other models are close in terms of accuracy with few exceptions. For instance, Albert shows a relatively large gap of 9\% for F1 scores, considering the condition $|R|\geq 0.9$ and a test ratio of 80\%. Distilbert performs generally well with a precision loss of few percentage points at most, compared to Roberta. In some cases, Distilbert performs even slightly better (e.g., $|R|\geq0.8$ with 80\% test ratio), cementing its status as a valid alternative to Roberta for the correlation prediction problem.

The following three tables (Table~\ref{tab:robdat} to \ref{tab:disdat}) report results for the scenario considered in Section~\ref{sec:scenarios}. They separate training and test data not at the granularity of column pairs. Instead, they separate them at the granularity of data sets. The following three tables report results for the three models (the semantics of all columns is the same as in the previous tables).

\begin{table}
    \centering
    \caption{Correlation prediction quality for data set separation and the Roberta transformer ($p\leq 0.05$).}
    \label{tab:robdat}
    \begin{tabular}{llllllll}
    \toprule[1pt]
    $\mathbf{|R|}$ & Test & F1 & Rec & Pre & Acc & MCC \\
    \midrule[1pt]
0.8 & 0.2 & 66 & 82 & 55 & 82 & 56 \\ \midrule
0.8 & 0.8 & 61 & 79 & 50 & 78 & 49 \\ \midrule
0.9 & 0.2 & 60 & 85 & 47 & 81 & 53 \\ \midrule
0.9 & 0.8 & 59 & 79 & 48 & 80 & 50 \\ \midrule
0.95 & 0.2 & 59 & 87 & 44 & 82 & 53 \\ \midrule
0.95 & 0.8 & 59 & 80 & 46 & 82 & 51 \\ \midrule
0.99 & 0.2 & 55 & 89 & 40 & 83 & 52 \\ \midrule
0.99 & 0.8 & 56 & 87 & 41 & 81 & 51 \\ \bottomrule[1pt]
    \end{tabular}
\end{table}

\begin{table}
    \centering
    \caption{Correlation prediction quality for data set separation and the Albert transformer ($p\leq 0.05$).}
    \label{tab:albdat}
    \begin{tabular}{llllllll}
    \toprule[1pt]
    $\mathbf{|R|}$ & Test & F1 & Rec & Pre & Acc & MCC \\
    \midrule[1pt]
0.8 & 0.2 & 64 & 76 & 54 & 82 & 53 \\ \midrule
0.8 & 0.8 & 59 & 73 & 49 & 77 & 46 \\ \midrule
0.9 & 0.2 & 62 & 80 & 51 & 84 & 54 \\ \midrule
0.9 & 0.8 & 59 & 72 & 49 & 81 & 48 \\ \midrule
0.95 & 0.2 & 57 & 83 & 44 & 82 & 51 \\ \midrule
0.95 & 0.8 & 58 & 76 & 47 & 82 & 50 \\ \midrule
0.99 & 0.2 & 55 & 85 & 41 & 84 & 52 \\ \midrule
0.99 & 0.8 & 56 & 84 & 42 & 83 & 51 \\
\bottomrule[1pt]
    \end{tabular}
\end{table}

\begin{table}
    \centering
    \caption{Correlation prediction quality for data set separation and the Distilbert transformer ($p\leq 0.05$).}
    \label{tab:disdat}
    \begin{tabular}{llllllll}
    \toprule[1pt]
    $\mathbf{|R|}$ & Test & F1 & Rec & Pre & Acc & MCC \\
    \midrule[1pt]
0.8 & 0.2 & 66 & 81 & 55 & 83 & 56 \\ \midrule
0.8 & 0.8 & 61 & 76 & 51 & 78 & 48 \\ \midrule
0.9 & 0.2 & 63 & 82 & 51 & 84 & 55 \\ \midrule
0.9 & 0.8 & 60 & 77 & 49 & 81 & 50 \\ \midrule
0.95 & 0.2 & 59 & 87 & 44 & 82 & 53 \\ \midrule
0.95 & 0.8 & 59 & 81 & 46 & 81 & 51 \\ \midrule
0.99 & 0.2 & 56 & 89 & 41 & 83 & 53 \\ \midrule
0.99 & 0.8 & 56 & 84 & 42 & 82 & 51 \\
\bottomrule[1pt]
    \end{tabular}
\end{table}

For all models and metrics, accuracy decreases, compared to the previous partitioning methods. The relative ranking across models remains similar. All models benefit again from additional training data.

\balance

\end{document}